\documentclass[twocolumn,aps,prb,epsf]{revtex4}

\usepackage{amsmath}
\usepackage{amssymb}
\usepackage{epsfig}
\usepackage{graphicx}%
\begin{document}

\title{Phonon effects in molecular transistors: Quantal and classical treatment} 
\author{A. Mitra, I. Aleiner and A. J. Millis\\Department of Physics, Columbia University\\
538 West 120th St, NY, NY 10027}
\date{\today}

\begin{abstract}
We present a comprehensive theoretical treatment 
of the effect of electron-phonon interactions on molecular transistors,
including both quantal and classical limits. We study both equilibrated and 
out of equilibrium phonons. We present detailed results for conductance, noise and 
phonon distribution in two regimes.
One involves temperatures
large  as compared to the rate of electronic transitions on and off the dot; in  this 
limit our approach yields classical rate equations, which are solved numerically
for a wide range of parameters. 
The other regime is that of  low temperatures and weak electron-phonon
coupling where a perturbative approximation in the Keldysh formulation can be
applied.  The interplay between the phonon-induced
renormalization of the density of states on the quantum dot and the
phonon-induced renormalization of the dot-lead coupling is found to be
important. Whether or not the phonons are able to equilibrate in a time rapid
compared to the transit time of an electron through the dot is found to affect
the conductance. Observable signatures of phonon equilibration are presented.
We also discuss the nature of the low-T to high-T crossover.
\end{abstract}


\maketitle

\section{Introduction}
 
In recent years it has become possible to fabricate devices in which the
active element is a very small organic molecule \cite{Organicdevice}. Such a
device may be thought of as a 'quantum dot': a structure weakly coupled to the
macroscopic charge reservoirs ('leads') and small enough that the discrete nature
of the energy levels on the dot is important. Quantum dots fabricated using
conventional semiconductor technology have been extensively studied
experimentally \cite{Quantumdotrefs} and theoretically \cite{Quantumdottheory}.\
However, the use of small molecules may lead to new physics. In particular,
as electrons are added or removed from a small molecule, both the \textit{shape} of
the molecule and its \textit{position} relative to the leads may be altered. 
The energies associated with these changes are
not small, and the time scales may be comparable to those related to the flow
of electrons into and out of the molecule. Interesting recent data
indicate that these effects may lead to observable structures in the
conductance spectra of the dot \cite{Ralph02,Park00,Ralph03}.

The shape change may be thought of as a coupling of electrons on the molecule
to phonon modes of the molecule, while the position change corresponds to phonon
dependent tunneling matrix elements. 
The subject of electron-phonon coupling in
quantum dots has received much theoretical attention
\cite{Glazman87,Wingreen89,Gogolin02,Alexandrov02,Balatsky02,McKenzie,Prokofiev03,Aji,Flensberg,
Fedorets}.
In an early important work Glazman and Shekhter obtained analytic
expressions for the transmission probability
through the dot under   
conditions that the phonons are always in equilibrium \cite{Glazman87} . 
Their results for the transmission go a long way towards describing the
behavior of the phonon-coupled system when one is far from resonance. However
their treatment neglects the phonon renormalization of the dot-lead coupling,  
and thus  
gives rise to a zero bias conductance at resonance that is smaller than the value predicted by the
Breit-Wigner formula (this situation was not the main interest of \cite{Glazman87}).
(These issues were also recently discussed by Flensberg \cite{Flensberg}). 
The lack of renormalization of the dot-lead coupling appears in the 
treatment
carried out by other authors as well \cite{Wingreen89,Alexandrov02,Balatsky02,McKenzie}, 
and in addition some authors assert 
(incorrectly, we believe) that phonon sidebands
may be observed even in the linear-response conductance
\cite{Balatsky02,McKenzie} by tuning the gate voltage. 

In this paper we revisit the problem of the phonon coupled dot. We present a comprehensive
formalism valid both in the classical and quantal limits which resolves the ambiguities
in the present literature. We also use this formalism to address    
new issues related to the behavior of this system 
under strongly non-equilibrium conditions in the quantal and classical regime. We further
study the nature of the quantal-classical crossover and the noise spectra.
An important feature of our results in the
quantum regime is that the conductance peak height at resonance is unchanged by
electron-phonon interactions. 

A recent paper by McCarthy et al. \cite{Prokofiev03} 
treats the problem of the phonon coupled 
dot in the high-T classical regime and for phonons that couple to the leads.
They present results for the regime of ``equilibrated'' phonons strongly
coupled to a heat bath. 
The high-T regime of our work is similar to that of reference \cite{Prokofiev03},
but we also study the physics of out of equilibrium phonons. 

Aji et al. \cite{Aji} have studied the model 
under off-resonant conditions when the conductance is very low
and the current is due to elastic/inelastic cotunneling. They study the 
phonon sidebands in the case of equilibrated and unequilibrated phonons, 
for the case of an electron coupled to a molecular vibrational mode (our
model) and also the case of a phonon dependent tunneling amplitude. 
In our analysis, the exact treatment of the leads automatically takes
into account cotunneling processes, while the perturbative approximation
restricts our analysis to details of the first phonon sideband. We 
show how thermal effects wash out cotunneling.

The paper is organized as follows. In Section II we describe the model (sub-section A),
the important model parameters (sub-section B) and 
develop a density matrix formalism (sub-section C) that allows
one to obtain the probability distribution
for various states on the dot under out of equilibrium conditions. 
In Section-III we apply the high temperature approximation in the density matrix formalism and 
derive the rate equations for the dot occupation probabilities in the sequential
tunneling regime (sub-section A). 
We use these rate equations to calculate average current (sub-section B) and the dc noise power 
(sub-section C) as functions of gate and source-drain voltage for two limiting cases: 
phonons equilibrated to
the external world independent of the electron occupation (phonon
equilibration fast compared to dwell time of electron on molecule), and 
phonons uncoupled from the
external world and responding only to on-dot electrons (phonon equilibration
slow compared to dwell time of electron on dot). We find that in the case
of phonons uncoupled from a bath, under certain bias conditions the phonon
distribution can deviate far from equilibrium. 

Section-IV deals with the low temperature quantal regime of the phonon coupled dot 
where we use the Keldysh Green's
function technique to calculate the dc-current and phonon distribution function 
to leading order in the ratio of 
(electron-phonon coupling)/(tunneling rate to the leads)
for  the two cases of slow and fast phonon equilibration rate. 
Section IV is split into sub-section A that re-introduces the problem, sub-section
B that derives expressions for the exact eigenstates in the absence
of electron-phonon coupling. These eigenstates form a convenient 
 basis for carrying out the perturbative
Keldysh calculation which is outlined in sub-section C. Sub-sections D and E present results
for I-V for the two cases of slow and fast phonon equilibration rates.

Finally Section V studies the crossover from 
low-T to high-T regimes and
section VI is a summary of our work and its relation
to already existing literature.

\section{Model, parameters and formalism}

\subsection{Model}

\begin{figure}[b]
\epsfxsize=2.0in \centerline{\epsffile{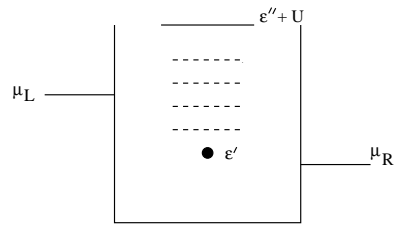}}
\vspace{0.5cm}
\caption{Energy level diagram. $\mu_{L,R}$
represent the chemical potential in the left and right leads respectively, while
$\epsilon^{\prime}$ represents the ground state of the singly occupied dot and 
the dashed lines indicate phonon excitations of the singly occupied dot.
The solid line indicates the energy $\epsilon^{\prime \prime} + U$ of the
doubly occupied dot. }
\label{fig:box}
\end{figure}

We consider the case of a molecule with a single level of degeneracy $d_g$
coupled to two leads, which we
label as 'left (L) ' and 'right (R)'. We suppose
that the electrons are coupled to two different kinds of phonons; an
internal vibrational mode of the
molecule of frequency $\omega_0$, which couples to the local charge, and 
a phonon mode labeled by a displacement operator $\hat{z}$,  that accounts
for the oscillations of the dot in an external confining potential of parabolicity $K$.
This phonon mode does not couple directly to the charge on the
dot, but results in an explicit $\hat{z}$ dependence in the left and right
tunneling matrix amplitudes $t_{i,k,\alpha}\{\hat{z}\}$. The full Hamiltonian is therefore,
\begin{equation}
H=H_{D}+H_{leads}+H_{t} \label{H}%
\end{equation}
with
\begin{eqnarray}
H_{D}  &=& \epsilon n_{d}+ \frac{U}{2} n_{d}\left(n_{d}-1\right)  +\lambda \omega_{0}\left(
b^{\dag}+b\right)  n_{d} \nonumber \\  
&+& \frac{p_z^2}{2M}  + \frac{1}{2} K z^2 
+ \omega_{0}b^{\dag}b \label{Hdot} \\
H_{leads} &=& \sum_{k,\alpha=L,R} \epsilon_k a^{\dagger}_{\alpha,k} a_{\alpha,k} 
\label{Hm}\\
H_{t}  &=&  \sum_{\alpha = L,R, k,\sigma,i=1,d_g} t_{i,k,\alpha}\{\hat{z}\} a^{\dagger}_{\alpha,k} d _{i,\sigma} + h.c  \label{Hmix}%
\end{eqnarray}
where $\sigma$ labels the spin index. Fig.~\ref{fig:box} shows the schematic of the energies 
considered.

Here the number of electrons on the molecule, $n_{d}$, is given by
\begin{equation}
n_{d}=\sum_{i=1,d_g,\sigma}d_{i,\sigma}^{\dag}d_{i,\sigma}, \label{nd}%
\end{equation}
and the parameter $U$ is the charging energy of the molecule.
We have defined the zero-phonon state of the vibrational mode of the molecule to
be the ground state when $n_{d}=0$, and we neglected anharmonicity in the
lattice part of the Hamiltonian (such anharmonicity is of course induced by
the electron-phonon coupling and an intrinsic anharmonicity could easily be
added). 

The dot-lead coupling $t_{i,k,\alpha}$
means that $[H,n_d] \neq 0$; in the absence of electron-phonon and 
many-body physics this implies an inverse lifetime for decay of an electron in
state $i$ on
the dot into a state of energy $\epsilon$ in lead $\alpha$,
\begin{equation}
\Gamma_{i,\alpha}(\epsilon) = 2 \pi \sum_{k} |t_{i,k,\alpha}(\hat{z} = 0)|^2 
\delta(\epsilon - \epsilon_{i,k})
\end{equation}

The density matrix $\rho$ of the full Hamiltonian $H$ obeys the equation of motion
\begin{equation}
\frac{d\rho(t)}{dt} = -i[H,\rho(t)]
\label{rho_eom}
\end{equation}
The current $I$  and the noise $S$ through the lead $\alpha$ are given by
\begin{eqnarray}
\langle I_{\alpha}(t) \rangle &=& Tr\hat{\rho}(t) \hat{I}_{\alpha} \label{defn}\\
S_{\alpha}(t) &=& Tr \hat{\rho}(t)\left(\delta I(t) \delta I(0) + \delta I(0) \delta I(t)  \right)
\end{eqnarray}
where $\delta I(t) = I(t) - \langle I \rangle$ and the current operator through the 
$\alpha$ lead is given by
\begin{eqnarray}
\hat{I}_{\alpha} &=& \frac{dN_{\alpha}}{dt} = i[H,N_{\alpha}] \\ \nonumber
&=& i \sum_{k,\sigma,i} \left ( t_{i,\alpha}(\hat{z}) {a^{\dagger}}_{\alpha,k,\sigma}d_i - h.c \right) 
\end{eqnarray}
$N_{\alpha} = \sum_{k,\sigma} a^{\dagger}_{k,\sigma} a_{k,\sigma}$ 
being the number operator for the $\alpha$ lead.

\subsection{Parameters and regimes}

\begin{figure}[b]
\epsfxsize=2.5in \centerline{\epsffile{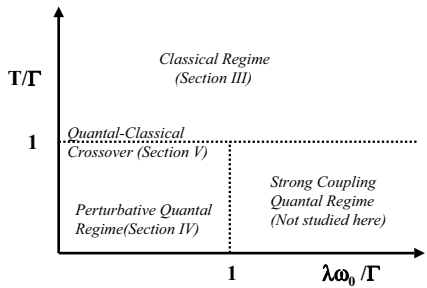}}
\vspace{0.5cm}
\caption{Overview of different regimes studied.}
\label{fig:over}
\end{figure}

The behaviour of the models we consider is specified by two important
dimensionless parameters: the  ratio of the temperature $T$ to a 
typical decay time $\Gamma$, and the product of the dimensionaless
coupling $\lambda$ and the ratio of the phonon frequency $\omega_0$ to $\Gamma$.
For large values of $T/\Gamma$ a classical rate equation analysis may be employed for 
all values of $\lambda \omega_0/\Gamma$; this is the subject of Section III.
For small values of $T/\Gamma$ a quantal treatment is required. Section IV reports
results for low T, obtained using perturbative calculations valid for 
$\lambda \omega_0/\Gamma \leq 1$, while Section V treats the quantal-classical 
crossover, also in the $\lambda \omega_0/\Gamma \leq 1$ regime. The quantal strong
coupling regime ($T/\Gamma < 1$,$\lambda \omega_0/\Gamma >1$) is a challenging
problem left for future reserach. The different regimes and the sections treating them
are shown in Fig.~\ref{fig:over}. 

\subsection{Formalism}
The essential assumption in the theory of molecular devices is that the leads are
in equilibrium independent of the state of the molecule. In order to implement this 
assumption it is often convenient to define a projected density matrix $\rho_s$
\begin{equation}
\rho_{s} = Tr_{leads}\{\rho(t)\} \otimes \rho_{leads}
\end{equation}
where $\rho_{leads}$ is the density matrix of the left and right leads that are in thermal equilibrium
at some specified chemical potentials ($\mu_L$ and $\mu_R$).
The density matrix of the dot (also known as the reduced density
matrix) is therefore given by  
\begin{equation}
\rho_D = Tr_{leads} \rho_s
\end{equation}
and for our model corresponds to projecting the full density matrix to the smaller
subspace of the  degrees of freedom of the dot electrons and the two types of phonon modes.

A complementary density matrix $\rho_t$ may also be defined such
that
\begin{equation}
\rho_t = \rho - \rho_s  
\end{equation}
from which it follows that  
\begin{equation}
Tr_{leads} \rho_t =0
\end{equation}
It is desirable to obtain the reduced density matrix of the dot $\rho_D$ because its
diagonal components relate directly to the occupation probabilities of the various 
states of the dot, and various expectation values (such as the current and noise)
may be expressed in terms of components of $\rho_D$.

In what follows we outline a general scheme for obtaining the reduced density 
matrix \cite{Zoller}.
The equation of motion for the full density matrix, 
Eq.~\ref{rho_eom}, implies
\begin{eqnarray}
\frac{d\rho_{sI}(t)}{dt}&=& \left( Tr_{leads} \frac{d\rho_I(t)}{dt}\right) \otimes\rho_{leads} \nonumber \\ 
&=& -i\left( Tr_{leads}[H_{tI}(t),\rho_{It}(t)] \right)\otimes \rho_{leads} \label{rhos_eom}\\ 
\frac{d\rho_{tI}(t)}{dt} &=& \frac{d\left( \rho_I(t)-\rho_{sI}(t)\right)}{dt} \nonumber \\ 
&=& - i[H_{tI}(t),\rho_{sI}(t) + \rho_{tI}(t)]  \nonumber \\ 
&+& i\left( Tr_{leads}[H_{tI},\rho_{tI}(t)] \right)\otimes \rho_{leads} \label{rhot_eom}
\end{eqnarray}
We have used the interaction representation defined by 
$\hat{O}_{I}(t) = e^{i(H_D + H_{leads})t} \hat{O}(t) e^{-i(H_D + H_{leads})t}$. 

In specifying the solution of Eqns.~\ref{rhos_eom} and \ref{rhot_eom}, we require an initial
condition. There are two common choices:(i) at the initial time, $\hat{\rho}$ corresponds 
to an equilibrium ensemble for $H$ with $\mu_L = \mu_R$, (ii) at the initial time ($t_i$) the
dot and leads are decoupled ($H_t = 0$) so that $\rho(t_i) = \rho_D \otimes \rho_{leads}$
with $\rho_D$ and $\rho_{leads}$ the equilibrium density matrices corresponding to the
uncoupled problem. We shall  be interested in steady state, so we will take the initial time 
$t_i = -\infty$. Further we shall be interested in cases in which the leads are not strongly
affected by the presence of the dot. In this case the boundary condition
(ii) is most convenient, but we note that if orthogonality effects or Luttinger liquid 
renormalizations of tunneling amplitudes are important then this choice may be less convenient.

Choosing boundary condition (ii), 
Equation \ref{rhot_eom} being a linear equation in $\rho_{tI}(t)$ may be formally solved as
\begin{equation}
\rho_{tI}(t) = -i \int_{-\infty}^{\infty} dt^{\prime} K_I(t,t^{\prime}) 
[H_{tI}(t^{\prime}), \rho_{sI}(t^{\prime})] 
\label{rhot2}
\end{equation}
where $K_I(t,t^{\prime})$ obeys the following operator equation,
\begin{eqnarray}
&\frac{dK_{I}(t,t^{\prime}) \bullet}{dt} + i[H_{tI},K_{I}(t,t^{\prime}) \bullet] 
\nonumber \\
&-i\left( Tr_{leads}[H_{tI},K_{I}(t,t^{\prime})\bullet] \right) 
\otimes \rho_{leads} 
= \delta(t-t^{\prime}) 
\label{Kdef}
\end{eqnarray}
where the symbol $\bullet$ denotes the operators acted on by $K_I$.
The combination of the second and third terms on the l.h.s.~of Eq.~\ref{Kdef} correspond to processes in which 
the particle distribution in the leads differ from  the equilibrium one. When 
substituted back in Eq.~\ref{rhot2}, these terms produce correlations between the dot
and lead variables. Processes corresponding to these correlations 
have been discussed in detail by Scholler and co-workers \cite{Schon94}, however 
their importance for the present problem is unclear.

Substituting Eq.~\ref{rhot2} into Eq.~\ref{rhos_eom}
leads to
\begin{equation}
\frac{d\rho_{DI(t)}}{dt} = - Tr_{leads} \int_{-\infty}^{\infty} dt^{\prime} 
[H_{tI}(t), K_I(t,t^{\prime}) [H_{tI}(t^{\prime}),\rho_{sI}(t^{\prime})]]
\label{rhos2}
\end{equation}
Eq.~\ref{rhos2} is a generalized Master equation that under steady state 
conditions (setting the l.h.s.~of Eq.~\ref{rhos2} to zero) 
can be written in the following form,
\begin{equation}
0 = \sum_{j} P^j R_{j \rightarrow i} 
\end{equation}
where $\langle i|\rho_D |i\rangle = P^i $, the probability of being in the i-th dot state.
(Under steady state conditions off-diagonal elements of the density matrix 
vanishes except in the case of accidental degeneracies).
Conservation of probability requires $\sum_{i} R_{j \rightarrow i} = 0$, from which 
it follows that  
\begin{equation}
0 = \sum_{j \neq i} P^{j} R_{j \rightarrow i} - P^i \sum_{j \neq i} R_{i \rightarrow j} 
\label{rate_gen}
\end{equation}
so that the quantities $R_{j \rightarrow i}$  may be interpreted as the various in-scattering 
and out-scattering rates.

A formal expression for the current may also be derived by using Eq.~\ref{defn} and observing
that $Tr \hat{I} \rho_s = 0$, so that
\begin{eqnarray}
\langle I \rangle &=& Tr \hat{I} \rho_{t} \label{curr_gen} \\  \nonumber
&=& -i Tr \int_{-\infty}^{\infty} dt^{\prime} 
[I_{I}(t)K_I(t,t^{\prime})), H_{tI}(t^{\prime})] \rho_{DI}(t^{\prime})
\otimes \rho_{leads}
\end{eqnarray}

Eqns.~\ref{rate_gen} and \ref{curr_gen} are our basic results. 
Further analysis depends on the 
specifics of the system and of the approximation chosen. At temperature T, only times $t < 1/T$
are relevent, so at sufficiently high $T$, we may approximate $K_I$ by its short time
behavior,
\begin{equation} 
K_{I}(t, t^{\prime}) \rightarrow \theta(t-t^{\prime})
\label{highT2}
\end{equation}
Indeed making this substitution in Eq.~\ref{rhot2} leads to the following expression
for $\rho_t$ in the Heisenberg representation,
\begin{eqnarray}
\rho_t &= -i \int_{-\infty}^{t} dt^{\prime}\, e^{-i(H_D +H_{leads})(t-t^{\prime})}[H_t,\rho_s(t^{\prime})] \label{rhot_sol} \label{rhot3} \\ \nonumber
&e^{i(H_D+H_{leads})(t-t^{\prime})}
\end{eqnarray}
The above expression for $\rho_t$ when substituted in the first line of Eq.~\ref{curr_gen} leads to
the widely accepted Meir-Wingreen \cite{Wingreen94} expression for the 
current,
\begin{eqnarray}
I_L &= \int_{-\infty}^{t} dt^{\prime} \int d\epsilon \, N(\epsilon)
Im\{e^{i\epsilon(t - t^{\prime})} 
G^{<}(t,t^{\prime}) + f_{L}(\epsilon) \label{MW1} \\ \nonumber
& e^{i\epsilon (t - t^{\prime})} G^{R}(t,t^{\prime}) \}
\end{eqnarray}
where $N(\epsilon)$ is the density of states, and 
 the Green's functions are defined as 
\begin{eqnarray}
G^R(t,t^{\prime}) &=& -i\theta(t-t^{\prime}) 
\langle \{d(t) \hat{t}_{kL}\{\hat{z}(t)\},d^{\dagger}(t^{\prime}) 
\hat{t}^{\dagger}_{kL}\{\hat{z}(t^{\prime})\}\}_+\rangle \\
G^<(t,t^{\prime})&=& i\langle 
d^{\dagger}(t^{\prime}) \hat{t}_{kL}^{\dagger}\{\hat{z}(t^{\prime})\} d(t) 
\hat{t}_{kL}\{\hat{z}(t)\}\rangle 
\end{eqnarray}

In general the second and third terms in Eq.~\ref{Kdef} imply $K_{I}(t) = \theta(t) k(t)$
where $k \neq 1$, suggesting the Meir-Wingreen formula is incomplete.
However we suspect that in the present problem in which there is no backscattering 
and no ``excitonic'' interaction between dot and lead variables, correlations
corresponding to $i[H_{tI},.] -iTr_{leads}[H_{tI},.] \otimes \rho_{leads} \neq 0  $
(Eq.~\ref{Kdef})
are irrelevant: any such configuration simply propagates away from the lead and does
not return. Some supporting evidence for this argument is provided by the
exact diagonalization results of Section III; however the issue warrants further
investigation.

For the simpler case of left and right phonon independent tunneling amplitudes
the Green's functions in Eq.~\ref{MW1} are simply the dot d-electron correlator. 
Wingreen et al. \cite{Wingreen94} showed that in this scenario and for the case of 
left and right tunneling amplitudes that are
proportional to each other (but with non-trivial on-dot interactions), 
the current through a single resonant level simplifies to
\begin{equation}
I = \int \frac{d\epsilon}{2\pi} 
\frac{\Gamma_L(\epsilon) \Gamma_R(\epsilon)}{\Gamma_L(\epsilon) + \Gamma_R(\epsilon)} A(\epsilon)
\label{MW2}
\end{equation}
where $A(\epsilon) = i (G^R_d(\epsilon) - G^A_d(\epsilon))$, with 
$G^R_d(t,t^{\prime}) = -i \theta(t-t^\prime) \langle \{d(t) , d^{\dagger}(t^{\prime})\}_+ \rangle $.
and $\Gamma_{L/R}(\epsilon) = 2\pi \sum_{k} t^2_{k,L/R} \delta(\epsilon_{k} - \epsilon)$.

When the temperature is the highest energy scale in the problem, 
approximation of Eq.~\ref{highT2} becomes exact, and further 
 the Green's functions in Eq.~\ref{MW1} are replaced by their 
short time behaviour \cite{Rate_Eqn}. This shall be explicitly demonstrated 
in the next section where we will derive 
rate equations for the on-dot probability distribution within the
sequential tunneling regime. Following this, in Section IV we shall carry out the Meir-Wingreen
prescription for calculating the current for general temperatures.

\section{High-temperature approximation: Rate Equations}

\subsection{Formalism}

In this section we 
shall carry out a high-T analysis for the simple case where the only phonon
mode the electron degrees of freedom couple to is the on-dot vibrational mode. (The case of
the phonons coupled to leads has been discussed is some detail by McCarthy et al.
\cite{Prokofiev03} within the high temperature approximation). 

It is often convenient to  choose a representation which is diagonal in the dot degrees
of freedom. In the present model this is achieved via  
a standard \cite{Mahanch4} canonical transformation.
Defining 
$S=\lambda\left(  \sum_{i,\sigma}d_{i,\sigma}^{\dag}d_{i,\sigma}\right)  \left(
b^{\dag}-b\right)$
and transforming all operators $\mathit{O}$ via $e^{S}\mathit{O}e^{-S}$ leads
to a transformed Hamiltonian $H^{\prime}=H_{dot}^{\prime}+H_{t}^{\prime} + H_{leads}$
with
\begin{eqnarray}
H_{D}^{\prime}  &  =\varepsilon^{\prime}n_{d}+\omega_{0}\tilde{b}^{\dag}%
\tilde{b}+\frac{\tilde{U}}{2}n_{d}\left(  n_{d}-1\right) \label{Hprimedot}\\
H_{t}^{\prime}  &  =\sum_{a=L,R,i}t_{i,a} \sum_{p,\sigma}\left(  \widehat
{X}a_{ap\sigma}^{\dag}d_{i,\sigma}+H.c.\right)  \label{Hprimemix}%
\end{eqnarray}
where the transformed phonon operator $\tilde{b}=b-\lambda\sum_{i,\sigma}d_{i,\sigma
}^{\dag}d$ $_{i,\sigma}$, so that the  phonon ground state depends on the dot
occupancy.
Moreover
$\epsilon^{\prime}=\epsilon-\lambda^{2}\omega_{0}$ 
is the 'polaron shift' in the energy for adding one electron to the
molecule and the  interaction parameter $U$ is also renormalized, but 
as we shall focus here on $U \rightarrow \infty$ we do not
write the renormalization explicitly. The
crucial phonon renormalization of the electron-lead coupling is given by
\begin{equation}
\widehat{X}=\exp\left[  -\lambda\left(  \tilde{b}^{\dag}-\tilde{b}\right)  \right]
\label{Xdef}%
\end{equation}

The high-T approximation proceeds from Eq.~\ref{rhot3} by making the Markov approximation
which involves replacing
$\rho_s(t^{\prime})$ in the integrand in Eq.~\ref{rhot3}
by $\rho_s(t)$. 
After substituting for $\rho_t$ in Eq.~\ref{rhos_eom} it is also convenient
to replace $\int_{-\infty}^{t} = 
\frac{1}{2} \int_{-\infty}^{\infty}$, which amounts to absorbing any level shifts induced by
the dot-lead coupling into the bare values of the parameters. 
Following this we obtain  
\begin{eqnarray}
&\frac{d\rho_s(t)}{dt} = -i[H_D + H_{leads},\rho_s] - \frac{1}{2}\int_{-\infty}^{\infty} 
dt^{\prime}\, \label{rhos_sol}\\\nonumber 
&[H_t, e^{-i(H_D+H_{leads})(t-t^{\prime})}[H_t,\rho_s(t)]e^{i(H_D+H_{leads})(t-t^{\prime})}]
\end{eqnarray}

On the assumption that orthogonality effects may be neglected, we may formally
take the trace over the lead degrees of freedom
and in the process arrive at coupled equations of motion for the 
various occupation probabilities of the dot. We outline the calculation
in detail for one of the 4 terms that one gets on opening up the commutator 
in Eq.~\ref{rhos_sol}.
\begin{eqnarray}
&Tr_{leads}\frac{\left(d\rho_{leads}\otimes\rho_D\right)}{dt} \\ \nonumber
&= -\frac{1}{2}\int_{-\infty}^{\infty} dt^{\prime}\,Tr_{leads} H_t e^{-i(H_D+H_{leads})(t-t^{\prime})}
\\ \nonumber
&H_t \rho_{leads} \otimes \rho_{D} e^{i(H_D+H_{leads})(t-t^{\prime})} + \dots 
\end{eqnarray}
Using the following relations
\begin{align}
Tr_{leads}{\rho_{leads}} &= 1 \\ \nonumber 
Tr_{leads} \left( \rho_{leads} a_{\alpha,k_1}^{\dagger} a_{\beta,k_2} \right)
&= \delta_{\alpha,\beta} \delta_{k_1,k_2} f(\epsilon_k-\mu_{\alpha}) \\ \nonumber
Tr_{leads} \left ( \rho_{leads} a_{\alpha,k_1} a_{\beta,k_2}^{\dagger} \right )
&= \delta_{\alpha,\beta} \delta_{k_1,k_2} \left( 1-f(\epsilon_k-\mu_{\alpha})
\right)\nonumber
\end{align}
we obtain
\begin{align}
\frac{d\rho_D}{dt} &= -\frac{1}{2} \int_{-\infty}^{\infty} dt^{\prime} 
\sum_{k,\alpha = L,R,i,j,\sigma} t^2_{\alpha} f_{k,\alpha} e^{i\epsilon_{k,\alpha}
(t-t^{\prime})} d_{i\sigma} \label{rhoD_eom}\\ \nonumber
&X e^{-iH_d(t-t^{\prime})} 
d^{\dagger}_{j\sigma}X^{\dagger} \rho_D e^{iH_D(t-t^{\prime})} \\ \nonumber  
&+ \, t^2_{\alpha}(1-f_{k,\alpha}) e^{-i \epsilon_{k,\alpha} (t-t^{\prime})} 
\\\nonumber
&d^{\dagger}_{i\sigma}X^{\dagger} e^{-iH_D(t-t^{\prime})} d_{j\sigma} X \rho^D e^{i H_D (t-t^{\prime})}
+ \cdots
\end{align}
We may now identify the probability $P^n_m$ of the dot being in a state with n electrons and
m phonons as,
\begin{equation}
P^n_m =  \langle n,m |\rho_D | n,m \rangle 
\end{equation}
Note that while $\rho_D$ is always diagonal in the dot electron number, it may
be off-diagonal in the phonon number (due to the presence of the $X$ operators in 
Eq.~\ref{rhoD_eom}). However such terms are negligibly small in comparision to the
components of $\rho_D$ that are diagonal in phonon number, and are therefore neglected
in our analysis. That is also the reason why we have dropped the first term 
in Eq.~\ref{rhos_sol} in the next set of equations.

We are now in a position to write rate (Master) equations for the electron-
phonon joint probabilities, which take the form 
\begin{eqnarray}
\dot{P}_{q}^{n}  &  =\sum_{a,q^{\prime}}f_{a}\left(  \left(  q-q^{\prime
}\right)  \omega_{0}+U(n-1)\right) \Gamma_{q,q^{\prime}}^{a}%
P_{q^{\prime}}^{\left(  n-1\right)  } \label{rateeqs} \\ \nonumber
&  + \left(  1-f_{a}\left(  \left(  q^{\prime}-q\right)
\omega_{0}+Un\right)  \right) \Gamma_{q,q^{\prime}}^{a}%
P_{q^{\prime}}^{\left(  n+1\right)  }\\
&  -\left(  1-f_{a}\left(  \left(  q-q^{\prime}\right)
\omega_{0}+U (n-1)\right)  \right)  \Gamma_{q^{\prime},q}^{a}P_{q}%
^{n}\nonumber\\
&  -f_{a}\left(  \left(  q^{\prime}-q\right)  \omega
_{0}+U n \right)  \Gamma_{q^{\prime},q}^{a}P_{q}^{n}\nonumber
\end{eqnarray}
Note that in our notation the upper index 
in $P^n_q$ always refers to the electron number and the lower index
the phonon number, while  $f_{a}(x)$ is short form for the Fermi function evaluated at
$f(x + \epsilon^{\prime} - \mu_a)$, $\mu_a$ being the chemical potential of lead $a$.

Thus within the high-T approximation the rate for going from an $n$ electron and $q$
phonon state on the dot to an $n-1$ electron $q^{\prime}$ phonon state is 
$R^{n \rightarrow n-1}_{q \rightarrow q^{\prime}} =  \sum_{a=L,R} f_{a}\left(  \left(  q-q^{\prime
}\right)  \omega_{0}+U(n-1)\right) \Gamma_{q,q^{\prime}}^{a} $, where 
$\Gamma_{q^{\prime},q}^{a}$ represents the 
transition rate involving hopping an 
electron from the dot to lead $a$ and
changing the phonon occupancy from $q$ (measured relative to the ground state
of $H_{D}^{\prime}$ with occupancy $n$) to $q^{\prime}$ (measured relative
to the ground state of $H_{D}^{\prime}$ with occupancy $n-1$) and is equal to
the transition rate involving hopping an electron from the lead $a$ to the dot
and changing the phonon occupancy from $q$ (measured relative to the ground
state of $H_{D}^{\prime}$ with occupancy $n-1$) to $q^{\prime}$ (measured
relative to the ground state of $H_{D}^{\prime}$ with occupancy $n$). More
explicitly
\cite{Mahanch4}
\begin{equation}
\Gamma_{q^{\prime},q}^{a}=\Gamma_{a}\left\vert <q^{\prime}|X|q>\right\vert
^{2} \label{gqq'}%
\end{equation}
The matrix element can be computed by standard methods
\cite{Mahanch4}; its absolute value $|<q|X|q'>|^2 \equiv X_{qq'}^2 $ is symmetric 
under interchange of $q$ and $q'$  and is%
\begin{equation}
X_{q<q'}^2= \left\vert\sum_{k=0,q}\frac{\left(  -\lambda^{2}\right)  ^{k}\left(  q!q^{\prime
}!\right)  ^{1/2}\lambda^{|q-q^{\prime}|}e^{-\lambda^{2}/2}}{\left(  k\right)
!\left( (q-k\right)  !\left(  k+\left\vert q^{\prime
}-q\right\vert \right)  !}\right\vert ^{2} \label{X}%
\end{equation}

As interesting special cases, we write several lowest operators:%
\begin{eqnarray}
X_{0n}  &  = e^{-\lambda^{2}/2} \frac{\lambda^n}{\sqrt{n!}}\label{X0n} \\\
X_{11}  &  = \left(  1-\lambda^{2}\right)  e^{-\lambda^{2}/2}\label{X11}\\
X_{21}  &  =\sqrt{2}\lambda\left(  1-\frac{\lambda^{2}}{2}\right)
e^{-\lambda^{2}/2}\label{X21}\\
X_{22}  &  =\left(  1-2\lambda^{2}+\frac{\lambda^{4}}{2}\right)
e^{-\lambda^{2}/2} \label{X22}
\end{eqnarray}
Observe that for certain values of $\lambda$ some of the matrix elements
vanish. This unusual behavior is an interference phenomenon, which is
slightly obscured by the notation. A state which has $q$ phonons excited
above {\it the ground state of the system with $n=0$ electrons} is 
a superposition (with varying sign) of many multiphonon states, when viewed in the
basis which diagonalizes the $n=1$ electron problem, and therefore the
transition described by $X_{qq'}$ is really a superposition of
many different transitions, which for some values of $\lambda$ may
destructively interfere. In 
several recent papers \cite{Alexandrov02}
the phonon renormalization of the molecule-lead coupling
is apparently omitted, or treated in an average manner
which neglects the $q,q'$ dependent structure.

\subsection{I-V characteristics}

In this subsection we shall discuss the I-V characteristics obtained from the
solution of the high temperature rate equations for two extreme cases. One is a scenario
where the phonons are not coupled to a bath and their number changes only
when electrons hop on and off the dot. The second case is when the phonons are
strongly coupled to a bath, and are always forced to be in equilibrium.

From Eq.~\ref{curr_gen} we obtain 
\begin{equation}
\langle I \rangle = Tr \rho_t(t) I_L 
\end{equation}
with 
\begin{align}
\hat{I}_{\alpha=L/R} = i t_L\sum_{k} \left ( a^{\dagger}_{\alpha,k} d X - d^{\dagger} X^{\dagger} 
a_{{\alpha},k} \right)
\label{I_L}
\end{align}

Using Eq.~\ref{rhot_sol} for $\rho_t$ and following the same procedure of tracing
out the metal degrees of freedom, we arrive at the following expression
for the current through the lead $a$ in terms of the joint 
probability distribution functions,
\begin{eqnarray}
I_a&=\sum_{n,q,q^{\prime}}(2d_g -n) P_{q}^{n} f_{a}\left(  \left(  q^{\prime
}-q\right)  \omega_{0}+U n\right)  \Gamma_{q,q^{\prime}}^{a} \label{Ia} \\ \nonumber
& - (n+1) P_{q}^{n+1}\left(
1-f_{a}\left(  \left(  q-q^{\prime}\right)  \omega_{0}+Un\right)
\right)  \Gamma_{q^{\prime},q}^{a}  
\end{eqnarray}
where the sum on $n$ is from $0$ to $(2d_g-1)$, $2d_g$ being the  
maximum occupation of the dot.
\vspace{0.3cm}

As an aside consider the simple case of spinless electron 
with no coupling to phonons with a non-degenerate dot level of
energy $\epsilon_d$.
In this case the rate equations give
us the following probability for a singly occupied level,
\begin{equation}
P^1 = \frac{\Gamma_L f(\epsilon_d-\mu_L) + \Gamma_R f(\epsilon_d-\mu_R)}{\Gamma_L + \Gamma_R}
\end{equation}
while the current from Eq.~\ref{Ia} is simply 
\begin{equation}
I_L = \frac{\Gamma_L \Gamma_R}{\Gamma_L + \Gamma_R}\{f(\epsilon_d-\mu_L) - f(\epsilon_d-\mu_R)\}
\end{equation}
On comparing the above expression with the exact solution for the
current obtained by Wingreen {\it et al.} (Eq.~\ref{MW2}) we can explicitly
see that the high-T approximation corresponds to
assuming the spectral function is a delta-function.

We shall now discuss the opposite limit, of phonons equilibrated 
to an independent heat bath, assumed to be at the same temperature
as the leads.
To implement this we force the 
probability distributions on the right hand side of Eq.~\ref{rateeqs}
to have the phonon-equilibrium form $P^{n}_{q} = P^n e^{-q\omega_{0}/T}(1-e^{-\omega_{0}/T})$.
In the $U \rightarrow \infty$ limit this ansatz implies that the 
probability $P^0$ that the dot is empty is given by,
\begin{equation}
P^0 = \frac{\sum_{a,q,q^{\prime}}\Gamma^a_{q,q{\prime}} e^{-q\omega_{0}/T}
\overline{f}_{a,q,q^{\prime}}}
{\sum_{a,q,q^{\prime}}2 \Gamma^a_{q,q{\prime}} e^{-q^{\prime}\omega_{0}/T} f_{a,q,q^{\prime}} 
+ \Gamma^a_{q,q^{\prime}} e^{-q\omega_{0}/T}\overline{f}_{a,q,q^{\prime}}}
\end{equation}
where $\overline{f}_{a,q,q^{\prime}} = 
1-f_a\left(\left(q-q^{\prime}\right)\omega_{0}\right)$,
$f_{a,q,q^{\prime}}= 1-\overline{f}_{a,q,q^{\prime}} $ while,  
$P^1 = 1 - P^0$.

\begin{figure}
\epsfxsize=2.5in \centerline{\epsffile{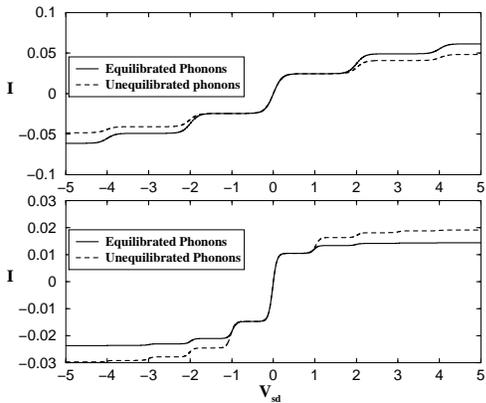}}
\caption{Current ($I$) vs source-drain voltage $V_{sd}$
for coupling constant $\lambda = 1.0$
$\omega_{0}=1$ and $T=0.05$. Upper panel is for $V_g=0.0$, while lower panel
is for $V_{g} = V_{sd}/2, \,\mu_R = 0$. $I$ is in units of $\frac{e T}{\hbar}$}
\label{fig:curr1}
\end{figure}

In general for both equilibrated and unequilibrated cases the rate equations 
may be written in the matrix form
\begin{equation}
{\dot{P}} = \hat{M} {P} 
\label{Pmatrix}
\end{equation}
Therefore under steady state conditions ($\dot{P}_n = 0$), the problem reduces
to finding the eigenvector corresponding to 
the zero eigenvalue of the matrix $\hat{M}$. 
We do this numerically. From these
solutions we have computed the current. Representative results are shown 
in Fig.~\ref{fig:curr1} which plots the low-T current as a function of $V_{sd}$ for two gate
voltages: $V_g = 0 \,\, (\mu_L= - \mu_R$, upper panel) 
and $V_g=\frac{V_{sd}}{2} \,\, (\mu_R=0$, lower panel), for both
equilibrated and unequilibrated phonons. (Note that the calculation is for
large values of $U$ which correspond to negligible double occupation probabilty
for the dot electrons). 

Steps (broadened by T) in the 
current associated with ``phonon side-bands'' are observed when the 
\underline{source-drain voltage} passes through an integer multiple 
of the phonon frequency. However, in the opposite
'linear response' limit $V_{sd} \rightarrow 0$ (not shown),
as $V_g$ is varied
we find just one main step in the $I-V$-curve,
as $V_g$ passes through 0, and only very tiny structures
(vanishing as $e^{-\omega_{0}/T}$, which is the probability of the dot being empty with
one phonon excited) when $V_g$ is a non-zero multiple of the 
phonon frequency. This result appears to differ from that stated by
other authors ~\cite{McKenzie,Balatsky02}
who find phonon side bands as $V_g$ is swept at
$V_{sd} \rightarrow 0$. The authors of ~\cite{McKenzie,Balatsky02} apparently 
neglected the fact that 

the phonon side-bands ``float'' {\it i.e.}, shift with the 
fermi level as  $V_g$  is changed.

Fig.~\ref{fig:curr1} reveals on first sight an apparently surprising result: 
for symmetric bias ($V_g=0$)
and for the coupling considered, the current is larger for equilibrated
phonons than for the unequilibrated case, whereas for the strongly asymmetric
case ($\mu_R=0$), the opposite is true. This is surprising because one expects
that in the unequilibrated case the phonons arrange themselves so as to maximize
the current.
To gain more insight into this phenomenon we have calculated the dependence of the
ratio of currents for unequilibrated and equilibrated phonons 
on the coupling $\lambda$ for different degrees of bias asymmetry.
We find that except for $\mu_R=0.0$ (the most asymmetric case) 
a minimum in the ratio occurs
for a $\lambda \sim 1$. This behaviour may be traced back to Eq.~\ref{X11},\ref{X22}
which reveal that higher order ``diagonal'' ($n$ phonon- $n$  phonon) matrix
elements vanish for a $\lambda \sim 1$. 

The steps in current may be conveniently parameterized by
the height (or the area, as the width is simply proportional
to $T$) of the corresponding peaks $G_{max}$ in the differential
conductance $G=dI/dV$.
Ratios of peak heights (or areas) provide a convenient experimental measure of
whether the phonons are in equilibrium.
At low T, the equilibrium phonon distribution
corresponds to occupancy only of the $n=0$ phonon state, so the n-th side band
involves a transition from the 0 phonon to the n phonon state. Therefore the ratios
of the peak heights or areas are controlled by ratios of $|X_{n0}|^2$. 
In particular 
Eqns.~\ref{X0n},~\ref{Ia} imply that if $\mu_L = -\mu_R$ and $T\ll \omega_{0}$,
\begin{equation}
\frac{G_{max}^n}{G_{max}^0}\Bigg\vert_{equil}
= \frac{|X_{n0}|^2}{2 |X_{00}|^2} = \frac{\lambda^{2n}}{2(n!)}
\end{equation}
Note that if $\mu_L \neq \mu_R$, then $\mu$ dependent changes in the 
occupation probabilities lead to additional, and not simply characterized
$n$ dependence.
\vspace{0.15cm}
\begin{figure}
\epsfxsize=2.5in \centerline{\epsffile{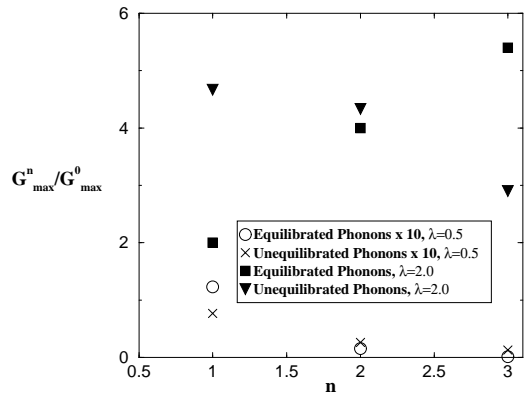}}
\vspace{-0.09cm}
\caption{
Ratio of differential conductance ($G$) peak heights for equilibrated and unequilibrated phonons and two 
different coupling strengths and $\mu_L = -\mu_R$. 
The points for $\lambda=0.5$ (open symbols) have been multiplied by 10.
\label{fig:G}
}
\end{figure}

Deviations from this pattern imply non-equilibrium phonons. As illustration we display
in Fig.~\ref{fig:G} $G_{max}$ values (normalized to the zero frequency peak) for equilibrium
and non-equilibrium phonons and a weak and strong electron phonon coupling. 
One sees that in the non-equilibrium case the peak heights display a non-systematic
dependence on electron-phonon coupling and peak index, but that in
general measurements of the $n=1$ and $n=2$ peaks reveal the effect clearly.
\begin{figure}
\epsfxsize=2.5in \centerline{\epsffile{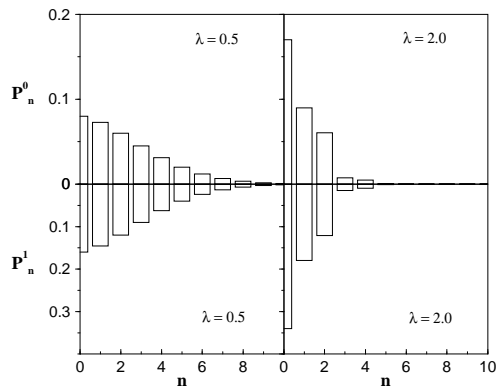}}
\caption{
Phonon probability distributions for two different electron-phonon coupling
constants calculated for $\mu_L = -\mu_R = 2\omega_{0}$.
\label{fig:pdist1}
}
\end{figure}

\begin{figure}
\epsfxsize=2.0in \centerline{\epsffile{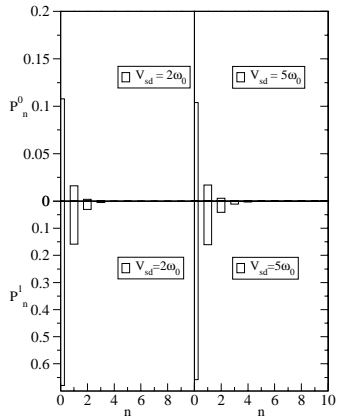}}
\caption{
Phonon probability distributions for weak electron-phonon coupling
($\lambda=0.5$) calculated for $\mu_L = 2\omega_{0}, \mu_R=0.0$ (left panel)
and $\mu_L = 5\omega_{0}, \mu_R=0.0$ (right panel). 
Note the saturation in the
probability distribution function which is also much closer to 
equilibrium. This situation is quite different for the same coupling
constant and symmetric bias (see Fig. \ref{fig:pdist1}).
\label{fig:pdist2}
}
\end{figure}

\begin{figure}
\epsfxsize=2.0in \centerline{\epsffile{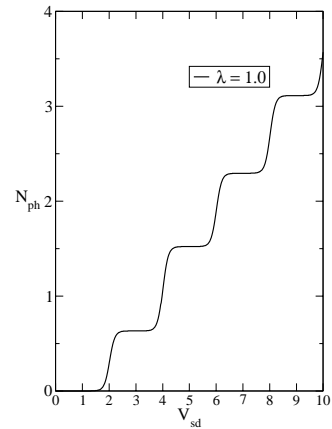}}
\caption{
Average phonon number under symmetric bias conditions, $\mu_L = - \mu_R$
, electron-phonon coupling $\lambda = 1.0$ and $\omega_0 = 20 T$.
\label{fig:ncl}
}
\end{figure}

It is also of interest to consider how far out of equilibrium the phonon distribution
may be driven. Fig.~\ref{fig:pdist1} shows the phonon occupation probabilities for  
weak and strong electron-phonon coupling and $V_g=0$. One sees
immediately that the phonon distribution function is farther from equilibrium 
for weak couplings than for strong couplings. We associate this effect to
the strong $\lambda$ dependence of operators $X_{0n}$, (Eq.~\ref{X0n}) which
allows the system at large $\lambda$ to ``jump down'' from a highly excited
state to one of low phonon occupancy. The deviation from equilibrium
is largest for $V_g=0$ for similar reasons. Fig.~\ref{fig:pdist2} illustrates the 
scenario of non-zero gate voltage or asymmetric bias conditions  $\mu_L = V_{sd}, \mu_R=0.0$.
Here the phonon distribution for weak coupling 
saturates with bias to a value which is closer to its equilibrium
distribution. As we shall show in Section IV, this gate voltage dependence of the non-equilibrium
phonon distribution function is recovered in the quantal regime $\Gamma \gg T$ as well.
Fig.~\ref{fig:ncl} is the average phonon number $N_{ph} = \sum_{n,m} n P^{m}_{n}$ for moderate
electron-phonon coupling. The steps in $N_{ph}-V_{sd}$ observed here coincide  
with the steps in I-V and
correspond to sequential (direct) tunneling. This is to be contrasted with 
the quantal regime (Section IV, Fig.~\ref{fig:Nph}) where $N_{ph}$ increases
continously with bias due to higher order cotunneling processes.

\subsection{DC Noise characteristics} 

Another important spectroscopic tool that is sensitive to the details
of the electron-phonon coupling and to the phonons distribution
is the current noise. In this sub-section we
outline the calculation of the dc current noise within the high-T 
approximation.
Quite generally, the current noise $S_{LL}(t)$ through the left lead 
is given by the following correlation function \cite{Zoller,Korotkov}
\begin{equation}
S_{LL}(t) = \frac{1}{2} Tr\, \rho \{ I_L(t)I_L(0) + I_L(0) I_L(t) \} - \left(Tr\, \rho I_L \right)^2
\label{noise1}
\end{equation}
Using the fact that $I_L(t) = e^{i H t} I_L(0) e^{-i H t}$, the above expression
for noise may be rewritten as 
\begin{equation}
S_{LL}(t) = \frac{1}{2} Tr\, \lambda(t) I_L(0)
\label{noise2}
\end{equation}
where
\begin{equation}
\lambda(t) = e^{-i H t} \{ \left ( I_L - \langle I_L \rangle \right)\rho + 
\rho \left( I_L - \langle I_L \rangle \right) \} e^{i H t}
\label{lambda}
\end{equation}
Since we are calculating the correlation of the same physical quantity
at two different times, we expect $S(t) = S(-t)$. Therefore we shall 
explicitly calculate $S(t)$ for positive times, for which we need to
calculate the causal function $\lambda(t)$ that obeys the equation of motion
\begin{eqnarray}
\frac{d\lambda(t)}{dt} &=& -i [H,\lambda(t)]+\delta(t)\{ \delta I_L 
\rho + \rho \delta I_L \}, t \geq 0
\label{lam_eom}  \\ \nonumber
&=& 0, \,\,\, t< 0
\end{eqnarray}
where $\delta I_L = I_L - \langle I_L \rangle $.
Note that  $Tr\, \lambda(0) = 0$, and we expect $\lambda(t)$ to be traceless 
at all times. Also from charge conservation it follows that the noise
across the left and right leads are equal ($S_{LL}(t) = S_{RR}(t)$).

We now solve the equation of motion for $\lambda$ by decomposing 
$\lambda= \lambda_{s} + \lambda_{t}$, where 
$\lambda_s = \lambda_{D} \otimes\lambda_{leads}$ is diagonal in the dot and 
lead variables, while $\lambda_t$  is off-diagonal. 
After doing a similar decomposition for the density
matrices $\rho = \rho_s + \rho_t$,
the equation of motion for $\lambda$ (Eq.~\ref{lam_eom}) may be re-written 
in a manner similar to that done for the equations of motion for $\rho$
in the previous section
\begin{align}
\frac{d\lambda_s}{dt} &= -i [H_t,\lambda_t] + \delta(t) \left ( 
I \rho_t + \rho_t I  -2 \langle I \rangle \rho_s \right ) \label{lams_eom}\\
\frac{d\lambda_t}{dt} &= -i [H_D + H_{leads}, \lambda_t] - i [H_t,\lambda_s] 
\label{lamt_eom} \\ \nonumber
&+ \delta(t) 
\left( I \rho_s  + \rho_s I - 2 \langle I \rangle \rho_t \right)
\end{align}
Note that we have dropped the  $[H_D + H_m,\lambda_s]$ term for the same reasons as
before, namely that the matrix elements of this term is off-diagonal in the phonon
number and is very small in comparision to the components of $\lambda_s$ that are
diagonal in phonon number.

The solution for $\lambda_t$ from Eq.~\ref{lamt_eom} is given by, 
\begin{align}
\lambda_t(t) &= -i \int_{-\infty}^{t} dt^{\prime}\, e^{-i(H_D+H_{leads})(t-t^{\prime})} 
[H_t,\lambda_s(t^{\prime})] \label{lamt_sol} \\ \nonumber
&e^{i(H_D+H_{leads})(t-t^{\prime})}
+ \theta(t)\, e^{-i(H_D+H_{leads})t} \\ \nonumber
&\left(I(0) \rho_s(0) + 
\rho_s(0) I(0) - 2 \langle I \rangle \rho_t(0) \right) 
e^{i (H_D+H_{leads})t} 
\end{align} 
Substituting Eq.~\ref{lamt_sol} in Eq.~\ref{lams_eom} and going through the
usual steps of extending the upper range of the integral to infinity, and
making the Markov approximation (which involves pulling $\lambda$ out
of the time integral), we arrive at the following matrix expression for the
equation of motion for $\lambda$,
\begin{eqnarray}
\frac{d\lambda_D}{dt} &=& \hat{M}\lambda_D + \delta(t) h , \, \, t \geq 0
\label{eom_lam} \\ \nonumber
& = & 0 , \, \,t<0
\end{eqnarray}
The matrix $\hat{M}$ is the same that enters in the equation of motion for $\rho_D$. In
arriving at the expression for the vector $h$, we have approximated oscillating factors such
as $e^{i \epsilon t} \sim 2 \pi \delta(t) \delta(\epsilon)$. 
Following this the vector $h$ has the structure
\begin{eqnarray}
h_q^n &= -2 \langle I_L \rangle P^n_q + 2 \sum_{q^{\prime},n}
(2d_g-n) P^{n}_{q^{\prime}} \label{h} \\ \nonumber 
&R^{L,n,n+1}_{q^{\prime},q} 
-(n + 1) P^{n+1}_{q^{\prime}} R^{L,n+1,n}_{q^{\prime},q}
\end{eqnarray}
where as before the sum on $n$ is from 0 to $2 d_g -1 $, $d_g$ being the number of
degenerate levels not counting spin.
Note that we are using the following short-hand notation,
\begin{eqnarray}
R^{a,n,n+1}_{q,q^{\prime}}= f_{a}\left(  \left(  q^{\prime
}-q\right)  \omega_{0}+U n\right)  \Gamma_{q,q^{\prime}}^{a} \label{Rdef1} \\
R^{a,n+1,n}_{q^{\prime},q}= (1-f_{a}\left(  \left(  q^{\prime
}-q\right)  \omega_{0}+U n\right))  \Gamma_{q,q^{\prime}}^{a} \label{Rdef2}
\end{eqnarray}
Moreover by using Eq.~\ref{Ia} it is easy to check that 
$h$ (whose components are given in Eq.~\ref{h}) is traceless.

Now we shall rewrite the Eq.~\ref{noise2} explicitly in terms of the
components of $\lambda_D$ and the 
steady state probabilities $ P^0_n $ and $P^1_n$ (components of 
reduced density
matrix $\rho_D$).
After some algebra one finds that
\begin{align}
S_{LL}(t) &= \frac{1}{2} Tr_{m,D} I_L \lambda_t, \,\, t \geq 0 \nonumber \\
&= - \langle I_L \rangle ^2 + \frac{1}{2}\left(S_{1L}(t) + S_{2L}(t)\right)
\label{SLL}
\end{align}
where $S_{LL}(-t) = S_{LL}(t)$,
and 
\begin{eqnarray}
S_{1L}(t) &= \sum_{q,q^{\prime},n} (2d_{g}-n) \lambda^{n}_{q}(t) 
R^{L,n,n+1}_{q,q^{\prime}} \label{S1L} \\ \nonumber
&-(n+1) \lambda^{n+1}_{q}(t) R^{L,n+1,n}_{q,q^{\prime}}
\end{eqnarray}
and
\begin{eqnarray}
S_{2L}(t) &= 2\delta(t) \sum_{q,q^{\prime},n} (2d_{g}-n) P^{n}_{q}(t) 
R^{L,n,n+1}_{q,q^{\prime}} \label{S2L} \\ \nonumber
&+(n+1) P^{n+1}_{q}(t) R^{L,n+1,n}_{q,q^{\prime}}
\end{eqnarray}
Note our notation $\lambda^n_q = \langle n,q |\lambda^D |n,q \rangle$ etc. 
where $\lambda^D = Tr_{leads} \lambda_s $. Also note that the $\delta(t)$ 
in the expression for $S_{2L}(t)$ again arises from replacing oscillating factors 
$e^{i \epsilon t} \sim 2 \pi \delta(t) \delta(\epsilon)$.

We find it convenient to perform the following shift of variables 
$\lambda^D \rightarrow \lambda^D + 2 \langle I_L \rangle P$. This shift of variables
does not affect the equations of motion Eq.~\ref{eom_lam} because $\hat{M} P = 0$
in steady state. However this shift of variables cancels the $\langle I_L \rangle ^2$
term in Eq.~\ref{SLL}.
\begin{figure}
\epsfxsize=2.2in \centerline{\epsffile{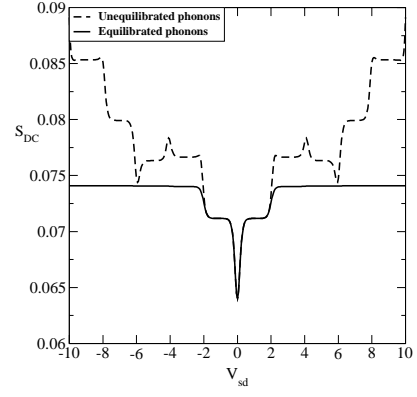}}
\vspace{1.3cm}
\caption{
DC Noise for $\lambda = 0.2$ and under symmetric bias conditions $\mu_L = -\mu_R = V_{sd}/2$.
$V_{sd}$ is in units of $\omega_0 = 20 T$. $S_{dc}$ is in units of $\frac{e^2 T}{\hbar}$.
\label{fig:S0.2}
}
\end{figure}

Collecting all the terms we arrive at the following expression
for the DC noise power 
\begin{equation}
\tilde{S}_{dc} = 2 \int_{-\infty}^{\infty} dt S_{LL}(t) 
\end{equation}
\begin{eqnarray}
\tilde{S}_{dc} &= 2\sum_{q,q^{\prime},n} (2d_{g}-n) (P^{n}_{q} + \tilde{\lambda}^{n}_{q}(0)) 
R^{L,n,n+1}_{q,q^{\prime}} \label{Sdc} \\ \nonumber
&-(n+1) (-P^{n+1}_{q} + \tilde{\lambda}^{n+1}_{q}(0) ) R^{L,n+1,n}_{q,q^{\prime}}
\end{eqnarray}
where the components of $\tilde{\lambda}(\omega=0)$ are obtained from solving the
matrix equation
\begin{equation}
\tilde{\lambda}^D(\omega=0) = - M^{-1} h
\end{equation}
Note that the matrix $\hat{M}$ has the property that $\sum_i M_{ij} = 0$,
so that one of its eigenvalues is $0$. However $h$ being
traceless, $M^{-1} h$ is well defined. 

Let us now look at the simple case of no electron-phonon coupling and 
spinless electrons. In that case the matrix $\hat{M}$ acqires the
simple $2\times2$ form (note that $f_a = 1/(1+e^{(\epsilon^{\prime} - \mu_a)/T})$
\begin{equation}
M  = \begin{pmatrix} -\left( \Gamma_L f_L + \Gamma_R f_R \right) & \Gamma_L (1- f_L) + 
\Gamma_R (1-f_R) \\ \Gamma_L f_L + \Gamma_R f_R  & -\left( \Gamma_L (1-f_L) + 
\Gamma_R (1-f_R) \right)
\label{M2by2}
\end{pmatrix}
\end{equation}
while the components of $\tilde{\lambda}({\omega=0})$ are given by
\begin{equation}
\tilde{\lambda}^0 = -\tilde{\lambda}^1 = -\frac{h^1}{\Gamma_L + \Gamma_R}
\end{equation}
where 
\begin{equation}
h^1 = 2 \Gamma_L P^0 f_L  - 2 \langle I_L \rangle   P^1
\end{equation}
The full expression for the dc-noise is
\begin{eqnarray}
\tilde{S}_{dc} &= 2\Gamma_L \left(P^0 f_L + P^1 (1-f_L ) \right)
-4\frac{\Gamma_L^2}{\Gamma_L + \Gamma_R} P^0 f_L \label{S_sim} \\ \nonumber
&+ 4\frac{\Gamma_L}{\Gamma_L + \Gamma_R}
\langle I_L \rangle P^1 
\end{eqnarray}
We may now derive expressions for the dc noise in two limits. The first one is
in the linear response regime $\mu_L = \mu_R$ so that $P^0 = 1 - f_L$
and $\langle I_L \rangle = 0$. In that case, 
\begin{equation}
\tilde{S}_{dc} =
4 \frac{\Gamma_L \Gamma_R}{\Gamma_L + \Gamma_R} f_L (1-f_L)
\end{equation}
which is the result expected from the fluctuation dissipation 
theorem $S_{dc} = 4 T G$. 

The other limit is $\mu_L = -\mu_R = eV/2 \gg T$ so that by using the
expression $P^1 = \frac{\Gamma_L f_L + \Gamma_R f_R}{\Gamma_L + \Gamma_R} $ 
and setting to zero combinations such as $f_L (1 - f_L)$
we obtain
\begin{equation}
\tilde{S}(\omega = 0) = 2 \frac{\Gamma_L^2 + \Gamma_R^2}{(\Gamma_L + \Gamma_R)^2}
\langle I_L \rangle 
\end{equation}
where $\langle I_L \rangle = \frac{\Gamma_L \Gamma_R}{\Gamma_L + \Gamma_R}$.
The above expression gives the standard shot noise result when 
$\Gamma_L \ll \Gamma_R$ ~\cite{Korotkov}   
\begin{figure}
\epsfxsize=2.2in \centerline{\epsffile{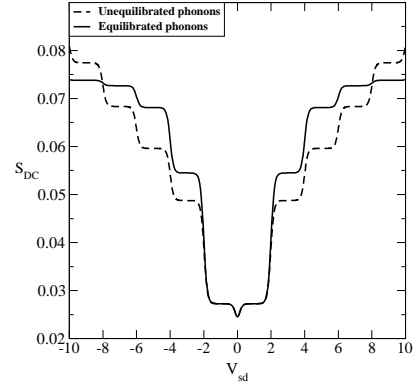}}
\vspace{1.3cm}
\caption{
DC Noise for $\lambda = 1.0$ and under symmetric bias conditions $\mu_L = -\mu_R = V_{sd}/2$.
$V_{sd}$ is in units of $\omega_0 = 20 T$. $S_{dc}$ is in units of $\frac{e^2 T}{\hbar}$.
\label{fig:S1}
}
\end{figure}

The results for the noise for the case of phonons in equilibrium and
the opposite case of phonons not coupled to any heat bath are illustrated
in figures \ref{fig:S0.2} (weak coupling) and \ref{fig:S1} (strong coupling). 
The difference between the equilibrated and unequilibrated
cases is more dramatic for smaller $\lambda$. For $\lambda = 0.2$ while only 
one phonon side band is  seen for the equilibrated case, very sharp 
phonon side bands are seen for the entire bias range
when the phonon distribution is far
out of equilibrium, with certain side bands appearing as peaks
(rather than steps) associated with the suppression of noise.

\section{Quantum theory of transport through a phonon-coupled dot}

\subsection{Overview}

In this section we present a fully quantum mechanical treatment of the
simple limit of the model considered in previous sections.
In order to carry out the calculation we restrict attention to a single
non-degenerate level with no onsite coulomb $U$, 
and to low orders in perturbation theory in the
electron-phonon coupling. The results shed light on the relation
between the Green's function formalism natural in the quantal treatment,
and the density matrix formalism natural in classical problems, and 
elucidate the quantal to classical crossover.
For the reader's convenience we reproduce here the limit of 
Eqns.~\ref{Hdot},\ref{Hm} and \ref{Hmix} which we study.
\begin{eqnarray}
H_{el} &=& \epsilon_0 d^{\dagger} d + 
\sum_{k,a=L,R} {\epsilon}_{k} a_{k,a}^{\dagger} a_{k,a} \label{H_in}\nonumber \\
&+&\sum_{k,a=L,R} t_{k,a} \left(d^{\dagger} a_{k,a} + h.c \right)\\
\\
H_{ph} &=& \omega_{0} b^{\dagger} b \label{H_ph}\\
H_{el-ph}&=& \lambda \omega_0 (b+b^{\dagger}) d^{\dagger} d \label{H_elph} 
\end{eqnarray}
Note that we have neglected the spin index, which may be restored
in the final expressions for the current by simply multiplying by a factor of 2.

We use two methods to analyze the above Hamiltonian. One is the Keldysh Green's
function method and the other is an explicit construction of the eigenstates
and thus the density matrix. We note that the Green's function method directly
computes the expectation value of operators at different times, bypassing the 
explicit construction of the density matrix. However equal time multiparticle Green's
functions correspond to moments of the density matrix, and permit in principle
its reconstruction.

The rest of the section is divided as follows. In subsection B we 
provide the exact solution 
in the absence of coupling to phonons.
The 
eigenstates of this system will form the basis for the perturbative calculations
that follow. In subsection C we outline the Keldysh calculation while in subsection
D we 
present results for the special case when the phonon distribution is
always in its ground state. In section E and F we generalize to the case when
the phonons are allowed to deviate from equilibrium and also 
supplement our results obtained from the Keldysh Green's function technique
by a perturbative calculation for  the phonon density matrix that allows us
to obtain the out of equilibrium phonon distribution function.

\subsection{Non-interacting dot and exact expressions for the
current for an interacting dot}

Following standard methods \cite{Mahanch4}, 
the exact eigenstates of $H_{el}$ (Eq.~\ref{H_in}) 
can be easily obtained.
The Hamiltonian after diagonalization is
\begin{equation}
H_{el} = \sum_{k,a=L,R} {\epsilon}_{k} \alpha_{k,a}^{\dagger} \alpha_{k,a} 
\label{hdmt}
\end{equation}
Note that while 
$a_{k,L/R}$ in Eq.~\ref{H_in} 
refer to states that live only on the left/right lead, 
an exact eigenstate of $H_{el}$ has non-vanishing amplitude in both leads. 
We write the exact solution in a scattering state basis in which 
$\alpha_{k,a=L/R}$
refers to a running wave incident
from the left/right lead with a certain amplitude of getting reflected back to the
starting lead and a corresponding amplitude for transmission to the other side.
The label L/R now refers to the lead from which the particle is incident, and therefore
determines the distribution function describing occupancy of the states.
The dot and metal electron creation/annihilation operators 
are related to these exact (scattering) eigenstates as follows
\begin{eqnarray}
a_{k,a} &=& \sum_{k^{\prime},b=L,R} \eta_{k a,k^{\prime} b} \alpha_{k^{\prime}, b}  \\
d &=& \sum_{k,a=L,R} \nu_{k,a} \alpha_{k,a}
\end{eqnarray}

The coefficients $\eta$ and $\nu$ in the above set of equations 
obtained from ensuring proper commutation relations are given by,
\begin{eqnarray}
\eta_{k a, k^{\prime} b} &=& \delta_{k,k^{\prime}} \delta_{a,b}  -\frac{t_{ka}\nu_{k^{\prime}b}}{{\epsilon}_{ka} - {\epsilon}_{k^{\prime} b} + i \delta } \\
\nu_{ka} &=& \frac{t_{ka}}{{\epsilon}_{ka} - \epsilon_0 - \sum_{k^{\prime} b} 
\frac{t^2_{k^{\prime}b}}{{\epsilon}_{ka} - {\epsilon}_{k^{\prime} b} - i \delta}}
\end{eqnarray}
We identify tunneling rates to the left and right leads by
\begin{equation}
\Gamma_{L/R}(\epsilon) = 2 \pi \sum_k t^2_{k,L/R} \delta(\epsilon-\epsilon_k) 
\end{equation}

The current through the left lead is given by 
\begin{eqnarray}
I_{L} &=&\frac{dN_L}{dt} =  -i\frac{e}{\hbar}\sum_{k} t_{kL} \left(d^{\dagger} a_{kL} - h.c. \right) \label{curr}\\ \nonumber
\langle I_{L} \rangle &=& 2 \frac{e}{\hbar}\sum_{k} t_{kL} Im \langle d^{\dagger} a_{kL} \rangle 	
\end{eqnarray}
and charge conservation requires $I_{L} = - I_{R}$. 
Plugging in the expressions for $d$ and $a_{kL}$  in terms of the exact eigenstates
$\alpha_{k a}$ into the expression for the current (Eq. \ref{curr}) we obtain  
\begin{equation}
\langle I_L \rangle = 2\frac{e}{\hbar} \sum_{k,k_1,k_2;\alpha,\beta=L,R} t_L Im \left(\eta_{k L, k_2 \beta} 
\nu^*_{k_1,\alpha}  \langle \alpha^{\dagger}_{k1, \alpha} \alpha_{k2,\beta}\rangle \right) 
\label{I_ex}
\end{equation}
In the non-interacting limit the expectation value of the exact eigenstates is
\begin{equation}
\langle \alpha^{\dagger}_{k_1 \alpha} \alpha_{k_2,\beta} \rangle  = \delta_{\alpha,\beta}
\delta_{k_1,k_2} f({\epsilon}_k-\mu_\beta)
\label{fermi}
\end{equation} 
where $f(x)$ is the Fermi-Dirac distribution function.
Substituting this in Eq. \ref{I_ex} we obtain 
\begin{equation}
\langle I_L \rangle = \frac{e}{\hbar}\int 
\frac{d{\epsilon}}{2 \pi} \frac{\Gamma_L(\epsilon) \Gamma_R(\epsilon)}{\Gamma_L(\epsilon) + \Gamma_R(\epsilon)} 
\{ f({\epsilon}-\mu_L) - f({\epsilon}-\mu_R) \} A({\epsilon})
\label{I_noneq}
\end{equation}
where $A({\epsilon})$ is the spectral density of the dot and is given by
\begin{equation}
A({\epsilon}) = \frac{\Gamma_L(\epsilon) + \Gamma_R(\epsilon)}{({\epsilon}-{\epsilon}_0 - \Sigma^{\prime}(\epsilon))^2 + (\frac{\Gamma(\epsilon)}{2})^2}
\end{equation}
Note that $\Sigma^{\prime}(\epsilon)$ is related to 
$\Gamma(\epsilon) = \Gamma_L(\epsilon) + \Gamma_R(\epsilon)$  by the 
usual Kramers-Kronig relation. 
For simplicity in our subsequent computation we assume energy independent density of
states and tunneling amplitude $t_{L,R}$, so that $\Sigma^{\prime} = 0$ and $\Gamma_{L,R}
$ are constants.
Eq.~\ref{I_noneq} agrees with the expression for the current that was derived 
by Wingreen {\it et al.} \cite{Wingreen94}
employing the Keldysh non-equilibrium technique. 

In the presence of electron-phonon interactions, 
the full Hamiltonian (now including Eq.~\ref{H_ph} and Eq.~\ref{H_elph})
takes the following form
in the basis of exact eigenstates of the non-interacting system,
\begin{eqnarray}
H &=& \sum_{k,a=L,R} {\epsilon}_{k} \alpha_{k,a}^{\dagger} \alpha_{k,a} + \omega_{0} b^{\dagger} b 
\label{HKel} \\
\nonumber
&+& \lambda \omega_0 (b+b^{\dagger})
\sum_{k,k^{\prime},a,b=L,R} \nu^{*}_{k,a}\nu_{k^{\prime},b} \alpha^{\dagger}_{ka}\alpha_{k^{\prime}b}  
\end{eqnarray}
A standard method for studying the nonequilibrium problem posed by $H$ is to define 
retarded Green's function for the electrons on the dot
\begin{eqnarray}
&G^{R}_d(t,t^{\prime}) = -i \theta(t-t^{\prime}) \langle \{d(t),d^{\dagger}(t^\prime)\}_+ \rangle \\
&= -i \theta(t - t^{\prime}) \sum_{ka,k^{\prime}b}\nu_{ka} \nu^{*}_{k^{\prime} b}
\langle \{\alpha_{ka}(t),\alpha^{\dagger}_{k^{\prime}b}(t^\prime)\}_+ \rangle
\end{eqnarray}
and the Keldysh Green's function for the dot electron,
\begin{eqnarray}
G^K_d(t,t^{\prime}) &=& -i \langle \{d(t),d^{\dagger}(t^\prime)\}_- \rangle \\
&=& -i  \sum_{ka,k^{\prime}b}\nu_{ka} \nu^{*}_{k^{\prime} b}
\langle \{\alpha_{ka}(t),\alpha^{\dagger}_{k^{\prime}b}(t^\prime)\}_- \rangle
\end{eqnarray}
Similarly for the phonons we define the corresponding retarded and Keldysh Green's functions, 
\begin{eqnarray}
D^R(t,t^{\prime}) &=& -i \theta(t-t^{\prime}) \langle \{b(t)+b^{\dagger}(t),b(t^{\prime}) + b^{\dagger}(t^\prime)\}_- \rangle  \label{dr_def}\\  
D^K(t,t^{\prime}) &=& -i  \langle \{b(t)+b^{\dagger}(t),b(t^{\prime})
+b^{\dagger}(t^\prime) 
\}_+ \rangle 
\label{dk_def}
\end{eqnarray}
Note that the phonon (electron) Keldysh
propagators in the equal time limit 
are $iD^K(t,t) = 
2(1 + 2 \langle b^{\dagger}b \rangle )$ ($iG^K_{d}(t,t) = 
1 - 2 \langle d^{\dagger}d \rangle  $) and are directly
related to the average
phonon (electron) number and therefore correspond to the first moments of
the density matrix. Higher order equal time correlators 
$\langle (b(t) b^{\dagger}(t))^n \rangle$ 
give higher moments of the density matrix, enabling in principle the full 
reduced density matrix $\rho_D$ to be
reconstructed.

The retarded dot Green's function for a single resonant level in the absence of phonons
can be easily obtained by using Eq.~\ref{fermi}
\begin{equation}
G^R_d(t,t^{\prime})= g^R(t_1,t_2) = \sum_{ka} |\nu_{ka}|^2  G^R_{ka,ka}(t_1,t_2)
\end{equation}
where $G^R_{k a,k a}(t,t^{\prime}) = -i \theta(t - t^{\prime}) \langle \{ \alpha_{ka}(t),
\alpha^{\dagger}_{k a}(t^\prime) \}_+ \rangle = -i \theta(t - t^{\prime}) e^{-i{\epsilon}_k(t-t^{\prime})}$.
It is now easy to see that in Fourier space the retarded Green's 
function for the non-interacting dot has the 
familiar form 
\begin{equation}
\tilde{g}^R(\omega) = \sum_{ka} \frac{|\nu_{ka}|^2}{\omega - {\epsilon}_k + i \delta} = \frac{1}{\omega -{\epsilon}_0 + i\frac{\Gamma}{2}}
\label{gr}
\end{equation}
In a similar manner the Keldysh Green's function in the absence of phonons 
is found to be 
\begin{eqnarray}
\tilde{g}^K(\omega) &=& 2 \pi i \sum_{ka} |\nu_{ka}|^2 \delta (\omega - e_k) 
\{ 2 f({\epsilon}_k-\mu_a) - 1 \} \label{gk} \\ \nonumber
&=& -i \frac{\Gamma_L \left(1-2f(\omega-\mu_L)\right) + \Gamma_R \left(1-2f(\omega-\mu_R)\right)}{(\omega-\epsilon_0)^2 + \frac{\Gamma^2}{4}}
\end{eqnarray}
where $f(x)= 1/(exp(\frac{x}{T}) + 1)$ denotes the Fermi distribution function.

Moreover in the non-interacting limit, the 
retarded and Keldysh phonon Green's functions ($D^{R/K}_0$) defined in 
Eq.~\ref{dr_def} and \ref{dk_def}
take the following form in Fourier space
\begin{eqnarray}
D^R_0(\omega) = \frac{2\omega_{0}}{\omega^2 - \omega_{0}^2 + i\delta sgn(\omega)} \label{dr}\\
D^K_0(\omega) = -2\pi i \{\delta(\omega + \omega_{0}) + \delta(\omega - 
\omega_{0}) \} \coth{\frac{\omega_0}{2T}}\label{dk}
\end{eqnarray}

In order to calculate the current for the case when electron-phonon interactions
are present, we shall use the result 
derived by Wingreen {\it et al.},
namely that the current through an interacting dot is still given by the expression
Eq.~\ref{MW2}, but with the spectral density 
$A(\epsilon) = 2i Im\{G^R_d(\epsilon)\}$, where $G^R_d$ is  the d-electron retarded 
Green's function calculated under 
appropriate nonequilibrium conditions and with respect to the full Hamiltonian H. 
We shall carry out this prescription for calculating the spectral density in the 
next section.

\subsection{Keldysh Greens function method: perturbative analysis}

In the presence of non-zero electron-phonon coupling,
the Dyson's equations we wish to solve may be 
written in the following compact form in $2 \times 2$ Keldysh 
space \cite{Keldysh_ref}
\begin{eqnarray}
\bf{G}_d^{-1} &=& \bf{g}_d^{-1} - \bf{{\tilde{\Sigma}}} \label{dys_e}\\ 
\bf{D}^{-1} &=& \bf{D}_0^{-1} - \bf{\Pi} \label{dys_p} 
\end{eqnarray}
where
\[ \bf{G}_d = 
\begin{pmatrix} 
 G^R_d & G^K_d  \\ 0 & G^A_d 
\end{pmatrix}
\]
is the local dot Green's function, 
and
\[ \bf{{\tilde{\Sigma}}} = 
\begin{pmatrix} 
 {\tilde{\Sigma}}^R & {\tilde{\Sigma}}^K  \\ 0 & {\tilde{\Sigma}}^A 
\end{pmatrix}
\]
is the electron self-energy due to electron-phonon interactions.
A similar matrix structure for the phonon propagator $\bf{D}$
and polarization $\bf{\Pi}$ in terms of retarded, advanced ($D^{R/A}$,$\Pi^{R,A}$)
and Keldysh ($D^{K}$,$\Pi^K$) components 
also exists. 
The non-interacting Green's functions
$\bf{g_d}$ and $\bf{D}_0$ have components that have been explicitly calculated in 
Eqn.~\ref{gr},~\ref{gk},~\ref{dr} and \ref{dk}. Note that the temperature enters
explicitly via the bare electron and phonon Green's function.

\begin{figure}
\epsfxsize=2.5in \centerline{\epsffile{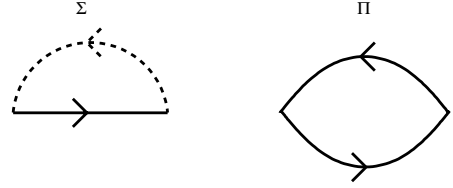}}
\vspace{0.01cm}
\caption{
Diagrams that correspond to the leading contribution to the electron
self-energy ($\Sigma$) and the phonon polarization ($\Pi$).
}
\label{fig:migdal}
\end{figure}

We analyze the equations perturbatively. The expansion parameter is
$\frac{\lambda \omega_0}{\Gamma}$,
and the leading non-trivial $\bf{\tilde{\Sigma}}$ and $\Pi$ are represented by the 
diagrams in Fig.~\ref{fig:migdal}. We write the perturbative expansion
in the usual self-energy language, but we note that in contrast to the
conventional band-electron case crossed diagrams for the electron Green's function
are not small relative to uncrossed diagrams, because the Green's function lacks
the pole structure found in the translation invariant case. Our results for
the electron Green's function and electron kinetic equation should be
understood to be perturbative in $\lambda$. 

To leading nontrivial  order 
it is sufficient to
calculate the phonon self-energy $\bf{\Pi}$ using the bare electron Green's function 
$\bf{g_d}$, but a correct calculation of $\bf{\tilde{\Sigma}}$ requires the use of the full 
$\bf{D}$.

The retarded phonon self energy (Fig. \ref{fig:migdal}) becomes
\begin{equation}
\Pi^R(t,t^{\prime}) = -\frac{i \lambda^2 \omega_0^2}{2} \{g^R(t,t^\prime) g^K(t^{\prime},t) 
+ g^K(t,t^{\prime}) g^A(t^{\prime},t)\} \label{Pir}
\end{equation}
while the Keldysh phonon self-energy is
\begin{eqnarray}
\Pi^K(t,t^{\prime}) &=  \frac{-i\lambda^2 \omega_0^2}{2}
\{g^R(t,t^{\prime})g^A(t^{\prime},t)  \\ \nonumber
& + g^A(t,t^{\prime})g^R(t^{\prime},t) + g^K(t,t^{\prime})g^K(t^{\prime},t) \}
\end{eqnarray}
Going into Fourier space and using Eqns.~\ref{gr} and \ref{gk} we obtain the following
expression for the real and imaginary parts of $\tilde{\Pi}^R(\omega)$ and $\tilde{\Pi}^K(\omega)$, 
\begin{eqnarray}
\tilde{\Pi}^R_{re}(\omega) &=& - \lambda^2 \{\Gamma_L T_1(\mu_L,\omega) + \Gamma_L T_1(\mu_L,-\omega) 
\label{Pirre} \\ \nonumber
&+& \Gamma_R T_1(\mu_R,\omega) + \Gamma_R T_1(\mu_R,-\omega) \} \\ 
\tilde{\Pi}^R_{im}(\omega) &=& i \lambda^2 \{ \Gamma_L T_2(\mu_L,\omega) - \Gamma_L T_2(\mu_L,-\omega) 
\label{Pirim} \\ \nonumber
&+& \Gamma_R T_2(\mu_R,\omega) - \Gamma_R T_2(\mu_R,-\omega) \}  
\end{eqnarray}
\begin{eqnarray}
&\frac{\Pi^K(\omega)}{2 i \lambda^2} = \frac{\Gamma_L^2}{2\Gamma} coth(\frac{\omega}{2T}) \left( T_2(\mu_L,\omega) - T_2(\mu_L,-\omega)\right) \label{Pik2} \\ \nonumber 
&+ \frac{\Gamma_R^2}{2\Gamma} coth(\frac{\omega}{2T}) \left( T_2(\mu_R,\omega) - T_2(\mu_R,-\omega)\right)\\ \nonumber  
&+ \frac{\Gamma_L \Gamma_R}{2\Gamma} coth(\frac{\omega + \mu_L - \mu_R}{2T}) 
\left( T_2(\mu_L,\omega) - T_2(\mu_R,-\omega)\right) \\ \nonumber
&+ \frac{\Gamma_L \Gamma_R}{2\Gamma} coth(\frac{\omega + \mu_R - \mu_L}{2T}) 
\left( T_2(\mu_R,\omega) - T_2(\mu_L,-\omega)\right) 
\end{eqnarray}
where we define the following integrals
\begin{equation}
T_1(\mu,\omega) =  \int \frac{d\omega_2}{2\pi}  
\frac{(\omega+\omega_2 - \epsilon_0) \{1-2f(\omega_2 - \mu)\}}
{((\omega_2 - \epsilon_0)^2 + \frac{\Gamma^2}{4})((\omega+\omega_2 - \epsilon_0)^2 + \frac{\Gamma^2}{4})}
\label{t1}
\end{equation}
and
\begin{equation}
T_2(\mu,\omega) = \frac{\Gamma}{2} \int \frac{d\omega_2}{2\pi} \frac{1-2f(\omega_2 - \mu)}
{((\omega_2 - \epsilon_0)^2 + \frac{\Gamma^2}{4})((\omega+\omega_2 - \epsilon_0)^2 + \frac{\Gamma^2}{4})}
\label{t2}
\end{equation}
Analytic expressions for $T_1(\mu,\omega)$ and $T_2(\mu,\omega)$ may be obtained at 
zero temperature, and are given in Appendix A.

Note that the combination $T_1(\mu_L,\Gamma_L,\omega)+T_1(\mu_L,\Gamma_L,-\omega)$
is symmetric, while  $T_2(\mu_L,\Gamma_L,\omega)-T_2(\mu_L,\Gamma_L,-\omega)$ is
asymmetric with respect to $\omega$. As a result, for all combination 
of couplings and
applied voltages
\begin{eqnarray}
\tilde{\Pi}^R_{re}(-\omega) = \tilde{\Pi}^R_{re}(\omega)\label{sym1}\\
\tilde{\Pi}^R_{im}(-\omega) = -\tilde{\Pi}^R_{im}(\omega)\label{sym2}\\
\tilde{\Pi}^K(-\omega) = \tilde{\Pi}^K(\omega)\label{sym3}
\end{eqnarray}

The retarded electron self-energy represented by the diagram in Fig.~\ref{fig:migdal} is
\begin{equation}
\Sigma^R(t,t^{\prime}) = \frac{i \lambda^2 \omega_0^2}{2}\{ g^R(t,t^{\prime}) D^K (t, t^{\prime}) + 
g^K(t,t^{\prime}) D^R(t,t^{\prime}) \}
\label{SR_def}
\end{equation}
Once we know the components of the polarization matrix $\bf{\Pi}$, we may use the Dyson equations
\ref{dys_p} together with the following parameterization for $D^K$ \cite{Vavilov_Aleiner} 
\begin{equation}
D^{K} = D^R \hat{f_k}^{ph} - \hat{f_k}^{ph} D^A \label{Dkph}
\end{equation}
to obtain,
\begin{equation}
D^R(\omega) = \frac{2\omega_0}{\omega^2 - \omega_0^2 + i\delta sgn(\omega) - 2 \omega_0 \Pi^R(\omega)}
\end{equation}
\begin{eqnarray}
\left(i\frac{\partial}{\partial t_1} + i\frac{\partial}{\partial t_2}\right)f_k^{ph}(t_1,t_2) 
&=& \Pi^R f_k^{ph}  \label{df_kph} \\ \nonumber
 &-& f_k^{ph} \Pi^A - \Pi^K
\end{eqnarray}
Note that our parameterization in terms of 
$f^{ph}_k$ is such that at equilibrium $f^{ph}_k(x) = \coth(x) = 1 + 2 n_B(x)$, 
$n_B$ being the Bose distribution function.

Besides $\Sigma^{R}$ it is also useful to evaluate the Keldysh component of
the self-energy $\Sigma^{K}$ that we will later use in order to calculate 
the distribution function of the dot. To the perturbative
order to which we work,
$\Sigma^K$ is related to the non-interacting
Green's function of the dot and the phonon Green's function as follows
\begin{eqnarray}
2\frac{\Sigma^K(t,t^{\prime})}{i \lambda^2 \omega_0^2} &=& 
g^R(t,t^{\prime})D^R(t,t^{\prime}) 
\label{SK_def} \\ \nonumber
&+& g^A(t,t^{\prime})D^A(t,t^{\prime}) + g^K(t,t^{\prime})D^K(t,t^{\prime}) 
\end{eqnarray}

Now that we know the retarded electron self-energy,
we can use Eq.~\ref{I_noneq} to calculate the current which is given by,
\begin{eqnarray}
&I(\mu_L,\mu_R) = \frac{e^2}{\hbar}\left(\frac{\Gamma_L \Gamma_R}{\Gamma_L + \Gamma_R}\right) 
\int \frac{d\omega}{2\pi} \left( f(\omega - \mu_L) 
- f(\omega - \mu_R) \right) \nonumber \\ 
&\frac{\Gamma - 2 \tilde {\Sigma}^R_{im}(\omega,\mu_L,\mu_R)}{(\omega - \epsilon_0 - \tilde{\Sigma}^R_{re}(\omega,\mu_L,\mu_R))^2 + 
\left(\frac{\Gamma}{2} - \tilde{\Sigma}^R_{im}(\omega,\mu_L,\mu_R) \right)^2}
\label{I_eq_K}
\end{eqnarray}
In the next two subsections we present our results for the two extreme cases 
of phonons strongly coupled to a heat bath, and therefore always
in equilibrium, and of phonons uncoupled to an external environment.

\subsection{Results : Equilibrated phonons}

\begin{figure}
\epsfxsize=2.0in \centerline{\epsffile{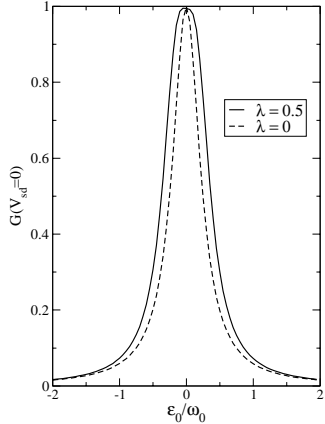}}
\vspace{0.01cm}
\caption{
Zero bias conductance (in units of $\frac{e^2}{h}$)
as a function of level energy (or gate voltage) with ($\omega_0 = 2 \Gamma$, solid line)
and without coupling to a vibrational mode (dotted line).  
\label{fig:GVg}
}
\end{figure}

In this sub-section we specialize to the case where the phonons are always in 
their ground state so that the phonon Green's functions $D^R$ and $D^K$ in
equations ~\ref{SR_def} and \ref{SK_def} are always calculated under the condition
that $\mu_L = \mu_R$, while all the non-equilibrium effects are included in the electron
Green's functions.
An important property of $\Sigma^R_{im}(\omega)$ is that at equilibrium it
is zero for $\omega= 0$, and from this and Eq.~\ref{I_eq_K}
it immediately
follows that the zero bias conductance even in the presence of electron-phonon
coupling has the form 
\begin{equation}
G(V=0) = \frac{e^2}{h}\frac{4 \Gamma_L \Gamma_R}{\left( \Gamma_L + \Gamma_R \right)^2}
\end{equation}
for the case where the two lead chemical potentials in addition to being equal
to each other are also aligned with the dot level $\epsilon_d$. Fig.~\ref{fig:GVg}
shows the gate voltage dependence of the zero bias conductance for symmetric broadening
and two values of $\lambda$.

\begin{figure}
\epsfxsize=2.5in \centerline{\epsffile{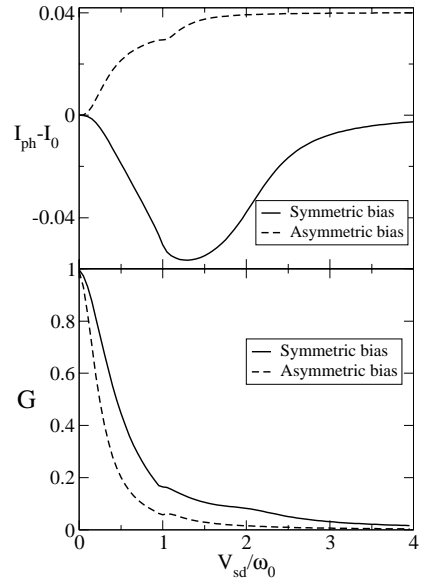}}
\vspace{0.01cm}
\caption{
Zero temperature current (in units of $\frac{e\Gamma}{h}$ and measured relative to 
no phonon current $I_0$ [upper panel] ) 
and conductance [lower panel] (in units of $\frac{e^2}{h}$)
under asymmetric ($\mu_R=0, \mu_L = V_{sd}$)
and symmetric ($\mu_L = - \mu_R = V_{sd}/2$) bias conditions and for equilibrated phonons.
The phonon frequency $\omega_0=2.0 \Gamma $ and $\Gamma_L = \Gamma_R = 0.5 \Gamma $. Moreover
the electron-phonon dimensionless coupling strength is $\lambda = 0.25$.
Note that under conditions of 
asymmetric bias, the current can saturate to a value larger than that for the device without
phonons ($\lambda = 0$).  
\label{fig:Ieq}
}
\end{figure}

Fig.~\ref{fig:Ieq} presents our result for the current (conductance) for the equilibrated phonon case at 
zero temperature. 
The top panel
is the difference between the current with and without electron-phonon coupling for two
different bias conditions. Note that for asymmetrically applied biases, the current with phonons
can take a value larger than that in the absence of phonons. 
This is due to a shift in the
center position of the spectral density (the total area under the spectral density being 
conserved). The lower panel is conductance for the same bias conditions. 
The conductance for the symmetric bias ($\mu_L = - \mu_R$ ) case shows two features, one 
at $V_{sd}/\omega_0 = 1.0$ and the other is a much broader feature at 
$V_{sd}/\omega_0 = 2.0$. While the former corresponds to the onset of in-elastic cotunneling, the
latter corresponds to the onset of sequential tunneling. (Under the asymmetric 
bias condition of $\mu_L = V_{sd}, \mu_R = 0$ one observes only sequential tunneling).

\begin{figure}
\epsfxsize=2.0in \centerline{\epsffile{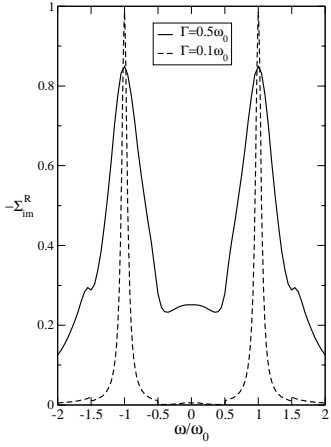}}
\vspace{0.01cm}
\caption{
Imaginary part of the electron self-energy (in units of $\Gamma$) due to phonons 
for the symmetric bias condition
$\mu_L = -\mu_R = V_{sd}/2  = \omega_0/2$, and $\lambda \omega_0 = 1 \Gamma$, level
energy $\epsilon_0 = 0$.
\label{fig:Sim}
}
\end{figure}

The transition from cotunneling dominant current to 
sequential tunneling dominant current can be understood by studying how the  
imaginary part of the electron self-energy due to interaction with phonons 
($\tilde{\Sigma}^R_{im}$) evolves with bias. Fig.~\ref{fig:Sim} 
shows $\tilde{\Sigma}^R_{im}(\omega)$ 
for the symmetric bias case $\mu_L = V_{sd}/2 = -\mu_R = \omega_0/2$ and for two different
values of $\Gamma/\omega_0$.
For simplicity we have considered the case where $\Gamma_L = \Gamma_R$. 
Under symmetric bias , $\tilde{\Sigma}^R_{im}$ increases rapidly for 
$|\omega| >  |\omega_0 - V_{sd}/2|$, while in calculating the current 
the spectral density is integrated from $-V_{sd}/2$ to $V_{sd}/2$ 
(see Eq.~\ref{I_eq_K}). 
Therefore clearly there is a threshold at $V_{sd} = \omega_0$ when 
the $\tilde{\Sigma}_{im}(\omega = \omega_0/2)$
jumps by  $\propto \frac{\Gamma}{\omega_0^2 + \Gamma^2}$, and this corresponds to the onset of
inelastic cotunneling. As the voltage is increased further, the range of integration
also increases to finally 
include the Lorentizian broadening centered around $\omega = \omega_0$, and
this corresponds to the onset of sequential tunneling. 

As $\Gamma$ is made smaller, the size of the step in $\tilde{\Sigma}^R_{im}(\omega = \omega_0/2)$
decreases, thus decreasing the cotunneling contribution to the conductance.
Moreover the lorentzian broadening in the self-energy
around $\omega_0$ also becomes narrower, so that the sequential tunneling 
peak in the conductance also gets sharper.

\begin{figure}
\epsfxsize=2.0in \centerline{\epsffile{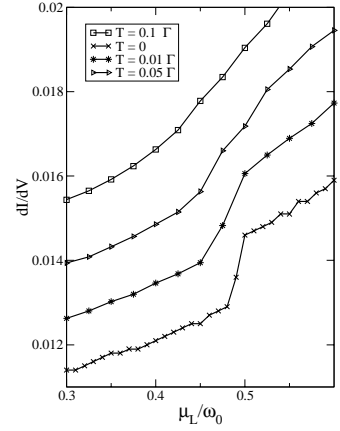}}
\vspace{0.01cm}
\caption{
Cotunneling dI/dV (in units of $\frac{e^2}{\hbar}$) for bias conditions 
$\mu_L = -\mu_R = V_{sd}/2$, level energy $\epsilon_0 = 4.0\Gamma$, $\omega_0 = 2 \Gamma$
, $\lambda \omega_0 = 1 \Gamma $ and for  phonons not
coupled to any heat bath.
\label{fig:cotun}
}
\end{figure}

Fig.~\ref{fig:cotun} shows the temperature dependence of inelastic cotunneling 
under conditions of no coupling to a heat bath (unequilibrated phonons).  
Calculations 
for this have been performed in the regime $\epsilon_0 > \mu_L, \mu_R$ so
that the current is entirely due to cotunneling. 
(The current under resonant 
conditions for this unequilibrated phonon case is discussed in detail in the next section).
It is clear from Fig.~\ref{fig:cotun} that inelastic cotunneling shows up as a 
step in $dI/dV$ that gets rounded very rapidly with increasing temperature.

\vspace{2.0cm}

\subsection{Results: Unequilibrated phonons}

\begin{figure}
\epsfxsize=2.5in \centerline{\epsffile{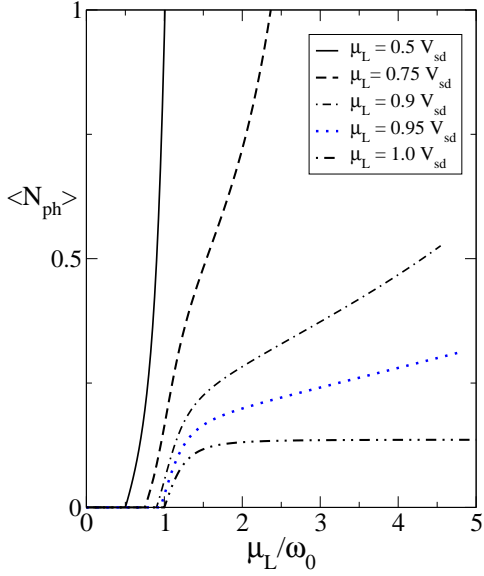}}
\vspace{1cm}
\caption{ Plot of average phonon occupation number for $\omega_0 = 2.0 \Gamma$, 
for several different bias conditions.
Except for the most asymmetric case ($\mu_L = V_{sd}$), the phonon number diverges for large enough 
bias voltages. Also note that the onset voltage for the deviation of the phonon
number from its equilibrium value (of zero) also continously shifts 
from $\mu_L = \omega_0$ (most asymmetric bias) to
$\mu_L = \omega_0/2$ (symmetric bias) signaling inelastic cotunneling.
\label{fig:Nph}
}
\end{figure}

\begin{figure}
\epsfxsize=2.5in \centerline{\epsffile{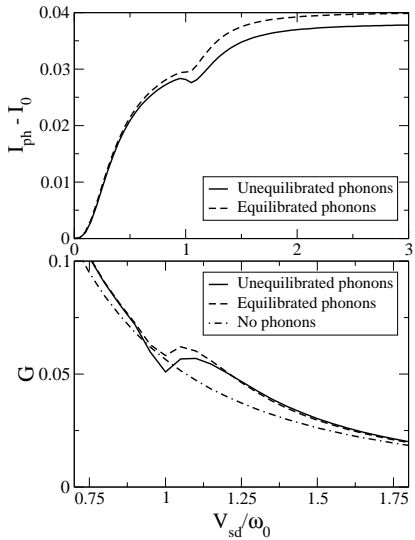}}
\vspace{0.01cm}
\caption{
Zero temperature conductance and current for equilibrated, unequilibrated and non-interacting 
levels under asymmetric bias conditions ($\mu_R=0, \mu_L = V_{sd}$). The phonon frequency 
$\omega_0=2.0 \Gamma $ and $\Gamma_L = \Gamma_R = 0.5 \Gamma $. Moreover
the electron-phonon dimensionless coupling strength is $\lambda = 0.25$. Note that 
$I$ (upper panel) is in units of $\frac{e\Gamma}{h}$ and $G$ (lower panel) is in units of 
$\frac{e^2}{h}$.
}
\label{fig:Ineq}
\end{figure}

For out of equilibrium conditions, 
we may derive a  quantum Boltzmann equation for the mean
phonon number, which for weak electron-phonon couplings is identified as 
$\langle N_{ph} \rangle = \frac{-1 + f^{ph}_k(\omega=\omega_0)}{2}$.
(This is from using Eq.~\ref{Dkph} and the fact that  
for weak couplings $D^R_{im}$ is almost a delta function at the phonon frequency). 
Therefore Eq.~\ref{df_kph} rewritten under steady state conditions and at an on-shell frequency
has the form
\begin{eqnarray}
0 &=& \langle N_{ph} \rangle \left(\Pi^K(\omega_0) + \Pi^R(\omega_0) - \Pi^A(\omega_0) \right) 
\label{qb_ph}\\ \nonumber
&-& (1 + \langle N_{ph} \rangle) \left( \Pi^K(\omega_0) -( \Pi^R(\omega_0) - \Pi^A(\omega_0))   
\right)
\end{eqnarray}
From this the phonon outscattering rate may be identified as 
$ \{\Pi^K(\omega_0) + \Pi^R(\omega_0) - \Pi^A(\omega_0) \} $, 
while the in-scattering rate may
be identified as $\{ \Pi^K(\omega_0) -( \Pi^R(\omega_0) - \Pi^A(\omega_0)) \}$.
$\langle N_{ph} \rangle$ is plotted in Fig.~\ref{fig:Nph} for a variety of bias conditions. 
The results here are similar to what was observed in the high-T classical calculation,
namely that for bias conditions under which the dot is half filled or close to it, the 
phonons tend to go far out of equilibrium. When the phonons deviate considerably
from their ground state, the corrections to the electron-self energy become
comparable to $\Gamma$ and one is no longer within the perturbative regime. 
Therefore the results for the current and conductance that we present here (Fig.~\ref{fig:Ineq}) are
for the case of $\mu_L = V_{sd}, \mu_R = 0.0$, when the phonons acquire a steady state 
distribution at large biases that is not far from its equilibrium value and a
perturbative approximation  is valid.

The upper panel of Fig.~\ref{fig:Ineq} plots the difference between the current with and 
without phonons for the asymmetric bias case, and for comparision this is plotted for
both equilibrated and unequilibrated phonons, while the lower panel is the
conductance peak corresponding to the first phonon
side-band. Again within perturbation theory 
the differences between these two cases is not significant.

\vspace{2cm}

\subsection{Perturbative calculation for Phonon distribution function}

So far the  out-of equilibrium Green's function technique allowed us to calculate various
averages such as current and mean phonon number. However 
it is also interesting to ask what the phonon probability distribution function itself is
under non-equilibrium steady state conditions. In order to do so we revert
to the density matrix formalism developed in Section II. 
The complication in calculating this quantity is the nontrivial form of the
operator equation Eq.~\ref{rhos2}.
One can however 
carry out the calculation for the rates to leading order in the electron-phonon
coupling, using the  
diagrammatic language developed
by various authors \cite{Schon94,Rammer}.  In implementing this we again find it 
convenient to be in the exact eigenstate basis for the non-interacting system (Eq.~\ref{HKel}).
The leading order contribution to the rates is obtained by expanding the exponentials
entering in the quantum rate equation ~\ref{rhos2} to leading order in the electron 
phonon coupling. The explicit form for the in-scattering and outscattering
rates is therefore
\begin{eqnarray} 
&R_{n \rightarrow n\pm 1} =  Tr_{leads} \int_{-\infty}^{0} dt 
\langle n |H_{e-ph}(0)| n \pm 1 \rangle \\ \nonumber
&\langle n \pm 1|H_{e-ph}(t) |n \rangle
\end{eqnarray}
where 
\begin{equation}
H_{e-ph}(t) =  \lambda \omega_0 (b(t) + b^{\dagger}(t))
\sum_{k,k^{\prime},a,b=L,R} \nu^{*}_{k,a}\nu_{k^{\prime},b} \alpha^{\dagger}_{ka}(t)
\alpha_{k^{\prime}b}(t)  
\end{equation}

\begin{figure}
\epsfxsize=2.4in \centerline{\epsffile{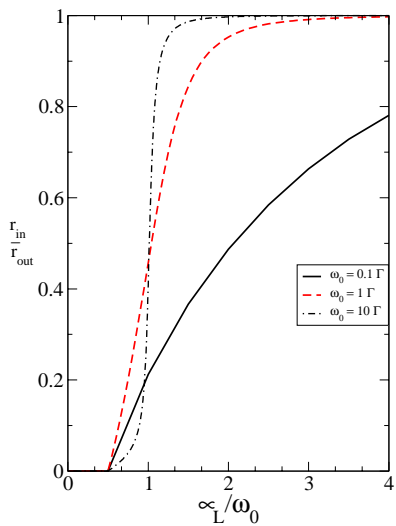}}
\caption{ Plot of ratio of phonon inscattering and outscattering rates 
$\frac{r_{in}}{r_{out}}$ for several different ratios of $\omega_0/\Gamma$.
The bias conditions are $\mu_L = -\mu_R$, and the onset of a non-zero $r_{in}$ at
$\mu_L = \omega_0/2$ is the sign of inelastic cotunneling. The high-T limit is approached 
for $\Gamma \ll \omega_0$, when the onset of nonzero $r_{in}$
shifts to the sequential tunneling limit of $\mu_L = \omega_0$. 
}
\label{fig:R}
\end{figure}

On evaluating the above expressions
we obtain the following quantum rate equation for the distribution function
\begin{eqnarray}
0 &=& -P^{ph}_n \left (R_{n \rightarrow n+1} + R_{n \rightarrow n-1} \right)  
 \label{qb_ph1} \\ \nonumber
&+& P^{ph}_{n+1} R_{n+1 \rightarrow n} +  P^{ph}_{n-1} R_{n-1 \rightarrow n}
\end{eqnarray}
where the in-scattering rate is given by  
\begin{eqnarray}
&R_{n\rightarrow n+1} = \lambda^2 \omega_0^2(n+1)\int\,\frac{d\epsilon}{4\pi} 
\label{R_in}\\ 
\nonumber
&\frac{\sum_{a,b=L,R} \Gamma_{a} \Gamma_{b} f(\epsilon-\mu_{a}) 
\left(1 - f(\epsilon - \omega_0 - \mu_{b})\right)}{\{(\epsilon - \epsilon_0)^2 + 
\frac{\Gamma^2}{4}\} \{(\epsilon - \epsilon_0 - \omega_0)^2 + \frac{\Gamma^2}{4}\}}
= (n+1) r_{in}  
\end{eqnarray}
and the out-scattering rate is given by
\begin{eqnarray}
&R_{n\rightarrow n-1} = \lambda^2 \omega_0^2n \int\,\frac{d\epsilon}{4\pi} 
\label{R_out}\\ \nonumber
&\frac{\sum_{a,b=L,R} \Gamma_{a} \Gamma_{b} f(\epsilon-\mu_{a}) 
\left(1 - f(\epsilon + \omega_0 - \mu_{b})\right)}{\{(\epsilon - \epsilon_0)^2 + 
\frac{\Gamma^2}{4}\} \{(\epsilon - \epsilon_0 + \omega_0)^2 + \frac{\Gamma^2}{4}\}}
= n r_{out} 
\end{eqnarray}
We have used short hand notations $r_{in/out}$ for the cumbersome integrals that
appear in the definition of $R_{n \rightarrow n+1}$ etc.
In terms of $r_{in/out}$ the quantum Boltzmann equation \ref{qb_ph1} takes the
form, 
\begin{equation}
P^{ph}_n \{ (n+1) \frac{r_{in}}{r_{out}} + n \} = P^{ph}_{n+1} (n+1) + (n) P^{ph}_{n-1}  
\frac{r_{in}}{r_{out}}       
\label{qpb2}
\end{equation}
It is easy to check that the above equation has a simple solution given by 
\begin{equation}
P^{ph}_n = (1 - \frac{r_{in}}{r_{out}}) \left(\frac{r_{in}}{r_{out}} \right)^n
\label{distsol}
\end{equation}
The quantity $\frac{r_{in}}{r_{out}}$ has been plotted for various combinations of
$\Gamma$ and $\omega_0$ in Fig.~\ref{fig:R}. The figure illustrates that the 
rapidity with which 
the phonon distribution diverges with bias depends on the relative sizes of 
$\Gamma$ and $\omega_0$. The stronger the coupling to the leads, the more easily
the phonons equilibrate.

In the high-T classical regime the rate equations for phonons for weak 
electron-phonon coupling has the same structure 
as Eq.~\ref{qpb2} but with modified scattering rates $r_{in/out}$. (This has been 
explicitly shown in Appendix B). It therefore follows that even in the high-T
regime, the phonon distribution function is given by Eq.~\ref{distsol}. 

Note that on comparing Eqns.~\ref{R_out},\ref{R_in} with 
Eqns.~\ref{Pirim},\ref{Pik2} we find, not surprisingly, 
\begin{equation}
\frac{r_{in}}{r_{out}} = \frac{\Pi^{K} - (\Pi^{R} - \Pi^A)}{\Pi^K + \Pi^R-\Pi^A}|_{\omega=\omega_0}
\end{equation}
where the polarization $\Pi$ were evaluated within the Keldysh Green's function approach.
Thus in particular 
$\langle N_{ph} \rangle = \frac{-D_K(t=0)+1}{2} = \sum_{n} n P^{ph}_n$, where $D_K$
has been calculated for phonons with no life-time broadening. 

\section{Low T to High-T crossover}

\begin{figure}
\epsfxsize=2.5in \centerline{\epsffile{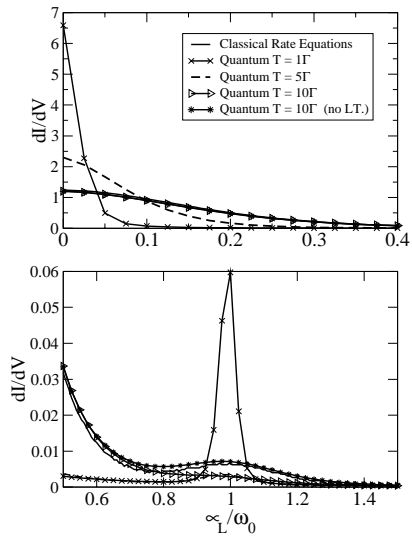}}
\vspace{0.01cm}
\caption{
Quantal-classical crossover effects in differential conductance spectra for
equilibrated phonons. Upper panel: temperature dependence of zero bias ``resonance''
peak computed as described in text and compared to results obtained from
classical rate equations (section III, but $U=0$). 
Lower panel: temperature dependence
of first phonon sideband, computed as described in text and compared to
results obtained from classical rate equations. The rapid thermal 
smearing of both central peak and phonon sideband is evident. Note that
in the phonon side-band case, broadening of the phonon level due
to electron phonon coupling leads to additional smearing not included 
in the rate equation model. The parameters are 
$\omega_0 = 100 \Gamma$, $\lambda = 0.1$ and $T= 1, 5, 10 \Gamma$. The bias
conditions are $\mu_L = -\mu_R$. Rate equation calculation: $\lambda = 0.1$, 
$T = 10 \Gamma$. 
Note: $dI/dV$ is in units of $\frac{e^2}{\hbar}\frac{\Gamma}{\omega_0}$.
\label{fig:Creq}
}
\end{figure}

\begin{figure}
\epsfxsize=2.0in \centerline{\epsffile{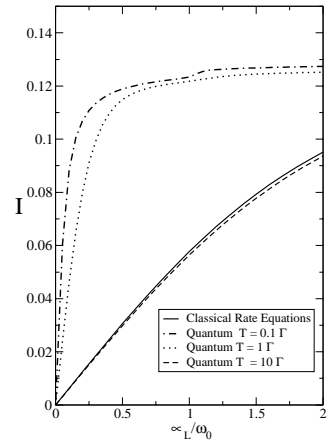}}
\vspace{0.01cm}
\caption{
Quantal to classical crossover effects on IV curve for unequilibrated phonons. 
The parameters are $\omega_0 = 10 \Gamma$, $\lambda = 0.05$ and the
temperature for the classical calculation corresponds to $T=10.0\Gamma$. The 
quantum calculation has been performed for $T= 0.1,1,10 \Gamma$. The results are
for asymmetric bias $\mu_R = 0.0, \mu_L = V_{sd}$ when the perturbation approximation
is valid. Note: $I$ is in units of $\frac{e \Gamma}{\hbar}$.
\label{fig:Crneq}
}
\end{figure}

We would now like to connect our low-T quantum calculation with the high-T 
calculation discussed in Section III. The high-T limit may
be reached by taking $T \gg \Gamma$ in the Keldysh calculation, and the
crossover from the $T=0$ to $T \gg \Gamma$ case has been illustrated in 
figures \ref{fig:Creq} and \ref{fig:Crneq} for equilibrated and
unequilibrated phonons respectively. 
The results for the equilibrated phonons have been presented for a rather
large electron-phonon coupling $\lambda \omega_0 = 10 \Gamma$ 
which strictly speaking is beyond the limits of validity of the perturbative
approximation used here,
in order to illustrate how the phonon sidebands
evolve with temperature.

The top panel of Fig.~\ref{fig:Creq} illustrates how the elastic resonance
broaden with temperature, while the lower panel shows the broadening of the
phonon side-band. 
As is evident from the two panels, at high temperatures the agreement between
the high-T rate equation calculation and the quantum calculation
is much better for the case of phonons with no life-time broadening. 
The effect of the life-time broadening due  to interactions with
electrons  is to round off the phonon side-band further.

Fig.~\ref{fig:Crneq} illustrates the cross-over from the low-T to high-T regime
under conditions of unequilibrated phonons and
within the perturbative limit of $\lambda \omega_0 = 0.5 \Gamma$ and asymmetric bias. 
The phonon side-bands vanish for $T > \lambda \omega_0$. 

\section{Conclusions}

\subsection{Summary}

In this paper we have studied a simple model of an electron-phonon 
coupled quantum dot, involving a (possibly degenerate) electron level
coupled to leads and to phonons. The problem has four different important
``internal'' parameters: a dimensionless electron-phonon coupling $\lambda$
(defined for example in Eq.~\ref{Hdot}), the ratio of the phonon frequency
$\omega_0$ to the broadening $\Gamma$ of the on-site level due to coupling
to the leads, the ratio of the temperature $T$ to the level width $\Gamma$. 
(A fourth parameter, the ratio of the rate $\gamma_{eq}$ at which the 
phonons relax to the heat bath characteristic of the device, to the mean electron 
current flow rate has only been studied in limiting cases).

The model admits two important sub-cases: of phonons coupled to the 
number of electrons on the molecule (Eq.~\ref{Hdot}), and of phonons coupled to
the dot-lead hybridization (Eq.~\ref{Hmix}).  Our formulation applies to
both cases, but we have focussed mostly on the former (McCarthy et al. 
have considered special features of the latter case in the classical regime).
In this paper we have attempted to present a general framework, within which
different special cases can be analyzed as desired. There are two important 
crossovers: the electronic quantal-classical crossover controlled by $T/\Gamma$,
and the phonon adiabatic/anti-adiabatic (phonon frequency long or short relative
to inverse electron dwell time on molecule) crossover controlled by $\omega_0/\Gamma$.
(The limit of $\omega_0 \ll T$ is not interesting).
In the classical limit (roughly, $T/\Gamma > 1$) this program has been carried out 
completely for all values of coupling $\lambda$ by us and by other workers. The relation 
between our results and those of other workers is discussed in detail in subsection B below.

The model has two external ``control'' parameters; the source-drain voltage difference
$V_{sd} = \mu_L - \mu_R$ (see Fig.~1) and the molecule one-electron addition energy 
$\epsilon^{\prime}$ measured relative to the average of the source-drain voltage 
$(\mu_L + \mu_R)/2$, and also referred to as the gate voltage. It is well known that the
conductance exhibits steps whenever one of the lead chemical potentials passes through  
the level energy $\epsilon^{\prime}$. The important new consequence of electron-phonon
coupling is the appearance of steps such as those shown in Fig.~\ref{fig:curr1} in the IV 
characteristics, when the source-drain voltage passes though an integer multiple of the 
phonon frequency. The existence of the phonon sidebands has been noted by several authors
\cite{Glazman87,Wingreen89}. An important new result (Fig.~\ref{fig:G}) of our work is the
systematic study of the dependence of the phonon heights on whether the phonon
distribution function is controlled by the nonequilibrium current or is 
equilibrated to a heat bath (Aji et al. presented similar information for the cotunneling
regime). We have also studied (Figs.~\ref{fig:pdist1},\ref{fig:pdist2},\ref{fig:Nph}) 
the non-equilibrium phonon distribution for different bias voltages. A further new result
of our work is the theory for noise in the classical limit (section III C) showing that
noise is a powerful spectroscopy of the degree to which the phonons are equilibrated 
(Fig.~\ref{fig:S0.2},~\ref{fig:S1}), especially in  the weak coupling limit.

In the quantal limit ($T < \Gamma $), our treatment is restricted to low orders of
perturbation theory in the coupling constant. Within this approximation we are
able to obtain results (Figs \ref{fig:cotun}, \ref{fig:Ineq}, \ref{fig:Creq} 
for the I-V characteristics including both the 
``direct tunneling'' ($\mu_L > \epsilon^{\prime} > \mu_R $ or conversely)
and ``cotunneling'' regimes ($\mu_L,\mu_R > \epsilon^{\prime} $ or conversely) and 
were able to treat the quantal to classical crossover (Figs \ref{fig:R},~\ref{fig:Creq},
\ref{fig:Crneq}). 
We presented both a 
diagrammatic (Keldysh) calculation and a solution based on the construction of
exact eigenstates, and confirmed that the peculiar broadening of the phonon
distribution found in the ``unequilibrated classical'' case survives also in the 
quantum limit.

\subsection{Other work}

In this section we consider the relation of our results to the extensive 
existing theoretical literature. The subject was pioneered by Glazman and
Shekter \cite{Glazman87}
who showed that the problem of a single electron transiting a resonant level
and coupled to phonons can be solved in complete analytic detail. 
They determined the form of the phonon side-bands in the transmission amplitude and showed
how the resonant behaviour was modified by the electron-phonon interaction.
Very similar results were subsequently obtained by Wingreen, Jacobsen and Wilkins 
\cite{Wingreen89}. However the single electron approximation used in
these papers is not applicable to the case of leads containing a Fermi sea of electrons.
The presence of other electrons blocks some of the intermediate states in the electron-
phonon scattering process, changing the form of the eigenstates and crucially blocks some of the
final states in the transmission process. This blocking ensures that the $T=0$ linear 
response conductance at resonance takes the quantized value if $2 e^2/h$,
whereas the on resonance transmission probabilities calculated in 
refs.~\cite{Glazman87, Wingreen89}
are less than unity. 
This issue was very recently also discussed by Flensberg \cite{Flensberg}.

One crucial consequence of the presence of Fermi seas in the leads is the ``floating''
of the phonon side-bands in the electron spectral function and thus in the IV curve. 
At $V_{sd}=0$, the features in the spectral function occur at energies displaced 
from the fermi-level by integer multiples of the phonon frequency, thus the corresponding
steps in the IV curve are observed when $V_{sd}$ is swept, but are not observed if
$V_g$ (i.e., the mean lead Fermi level) is changed at fixed small $V_{sd}$ 
(see Fig.~\ref{fig:GVg}).  Several authors 
\cite{McKenzie,Balatsky02,Alexandrov02} employ approximations to the 
electron Green's function corresponding roughly to those of 
Glazman and Shekhter or Wingreen and coworkers. 
These miss the above physics and erroneously predicts phonon side-bands as $V_g$ is varied 
\cite{McKenzie,Balatsky02}.

As also noted by Flensberg the approximations employed by refs.~\cite{McKenzie,Balatsky02}
amount to writing the dot Green's function as 
$G^R_d(t) = e^{(i\epsilon_0 - \Gamma)t} \langle X^{\dagger}(t)X(0)\rangle  $
(where the operators $X$ have been defined in Eq.~\ref{Xdef}).  This approximation
becomes exact in the high temperature limit, and can be used as the starting point of 
an alternative derivation of the rate equations. However some of the literature 
\cite{Alexandrov02} who have used this approach have treated the X-operator matrix elements in an 
approximate manner which does not capture the structure giving rise to the step
height variations displayed in our Fig.~\ref{fig:G}.

Finally Gogolin and Komnik have used an adiabatic (slow phonons, fast 
electrons) semiclassical approximation to explore the strongly electron-phonon
coupled regime. It is well known that at strong coupling the energy as a function
of phonon coordinate become bistable, signaling the onset of polaronic instability.
Ref.~\cite{Gogolin02} considered the behaviour of the polaronic state under non-equilibrium 
conditions, and observed that a bistable I-V characteristic could result. By contrast 
our rate equation analysis always leads to single valued I-V curves. The bistability corresponds
to a first-order ``energy landscape''; however the system under study 
is zero dimensional, and (at least within the approach used here) thermal and quantal
fluctuations allow the system to explore all of phase space wiping out any phase transition behaviour.
The calculation of Gogolin and Komnik are presented for a different regime ($T \ll \Gamma$),
but we suspect that fluctuations would also eliminate the apparent transition in that case.
An enhanced low frequency noise might however occur.

Four recent paper have appeared which present results consistent with those
presented here, but emphasizing somewhat different aspects of the physics. 
Mc Carthy et al.~\cite{Prokofiev03} used the rate equation approach of our section III,
our results reproduce theirs; however the main interest if ref ~\cite{Prokofiev03} 
was in the IV curves of phonons coupled to the dot-lead hybridization and the principal
focus was on thermally equilibrated phonons. Our result extend theirs by 
treating the nonequilibrium case (see eg. Fig.~\ref{fig:pdist1}) and the noise 
(see eg. Fig.~\ref{fig:S1}). 

Fedorets and coworkers \cite{Fedorets} have analyzed the same problem as
McCarthy et al. but in the limit of very weak coupling of phonons to a heat 
bath. They find that an instability occurs for $V_{sd}$ greater than a critical value. 
Interestingly in the weak coupling limit the critical $V_{sd}$ reported is identical 
to the critical $V_{sd}$ found in our calculation above which the phonon distribution 
become broad (see Fig.~\ref{fig:pdist1}). According to the refs.~\cite{Fedorets} the 
instability occurs if one has both lead coupled and dot coupled phonons, whereas the 
broad distribution is generic. However, this relation deserves further exploration. 
Here we observe that the calculations of refs.~\cite{Fedorets} (as well as those 
of our Section III) are based on a sequential-tunneling approximation. In this
approximation when the threshold is exceeded the distribution we find changes rapidly 
(in the weak electron-phonon coupling limit) from very narrow to very broad.
However if cotunneling processes (in particular the electron contribution to the 
phonon lifetime) are included, the transition becomes broadened with the onset 
moving to a lower voltage (see Fig.~\ref{fig:Nph}). We find the breadth of the 
distribution to depend strongly on the bias asymmetry , {\it{i.e.}}, to the average
occupation of the dot. 

Flensberg \cite{Flensberg} has used an approximate equation of motion approach to analyze the
quantum limit, determining in particular the equilibrium dot spectral function and 
presenting a clear analysis of the terms omitted in previous approaches \cite{Glazman87,Wingreen89}.
Where there is overlap our results are in agreement with his, however non-equilibrium conditions
have not been considered in this reference \cite{Flensberg}.

Finally Aji, Moore and Varma  ~\cite{Aji} have considered phonon effects on the cotunneling spectrum. 
Their results are in essential agreement with ours, however we note that
the steps in $dI/dV$ which they report to be vertical at $T=0$ are in fact smoothed by the
phonon lifetime arising from electron-phonon coupling (see Fig.~\ref{fig:cotun}) or to 
coupling to a heat bath. Also we find that the steps are thermally broadened into 
indetectability at a relatively low T fixed by the step height and the phonon lifetime induced
slope, rather than generically visible up to temperatures of order the phonon frequency, 
as stated in the reference \cite{Aji}.

As far as experiments are concerned, three recent papers ~\cite{Ralph02,Park00,Ralph03}
have observed signatures in the current-voltage characteristics which are attributed to 
coupling to a molecular vibrational mode. A direct comparison to our results is 
not yet feasible because noise measurements have not been performed, and
most of the samples studied show one or at most two phonon sidebands. (Only one of the
samples of Ref.~\cite{Park00} was reported to show more).

\subsection{Future Directions}

Finally we briefly mention a few issues raised by this work, but not resolved. One 
important research area is the extension of our quantum limit results beyond the 
perturbative regime (in particular to the nonequilibrium polaron limit), 
a second is to obtain the frequency dependence of the noise spectra in the classical
and quantal regimes, a third issue is the crossover between equilibrated and 
unequilibrated phonons, and a fourth issue is to explore the connection between the
nonequilibrium distributions we find and the instabilities discussed by the Chalmers
group \cite{Fedorets}. 
Another interesting question is to explore the effect of image charges induced in the surrounding electrodes 
on I-V ~\cite{Flensberg03} which have been argued to be important in certain experiments
\cite{Ralph03}.
Work in these directions is in progress.

\textit{Acknowledgements: }\ The authors would like to thank
L. Glazman and K. Matveev for helpful conversations.
We acknowledge support from NSF-DMR-0338376 and Columbia University, and AJM
thanks the ESPCI for hospitality while part of this work was completed.
This work was partially supported by the Nanoscale Science and Engineering
Initiative of the National Science Foundation under 
NSF Award Number CHE-0117752 and by the new York State Office
of Science, Technology and Academic Reserach (NYSTAR).

\appendix
\section{Zero temperature phonon polarization}

At zero temperature the function $1-2f(x) = sgn(x)$, and the integral 
Eq.~\ref{t1} and ~\ref{t2} may be performed explicitly. The corresponding
expressions are 
\begin{eqnarray}
&T_1(\mu_L,\omega) = \frac{1}{2 \pi (\omega^2 + \Gamma^2)}\ln{\frac{(\mu_L-\epsilon_0)^2 + 
\frac{\Gamma^2}{4}}{(\mu_L + \omega - \epsilon_0)^2 + \frac{\Gamma^2}{4}}} \label{t1_zeroT}\\ \nonumber
&+\frac{1}{2 \pi \omega }\left (\frac{2 \Gamma}{\omega^2 + \Gamma^2} \right)
\arctan{\frac{\mu_L+\omega - \epsilon_0}{\Gamma/2}}
\\ \nonumber
&- \frac{2}{\pi\Gamma\omega} \left (\frac{\frac{\Gamma^2}{2} + \omega^2}{\Gamma^2 + \omega^2}\right) 
\arctan{\frac{\mu_L - \epsilon_0}{\Gamma/2}}
\end{eqnarray}
and
\begin{eqnarray}
&T_2(\mu_L,\omega) = -\frac{1}{\pi (\omega^2 + \Gamma^2)}\left 
(\arctan{\frac{\mu_L+\omega -\epsilon_0}{\Gamma/2}} + 
\arctan{\frac{\mu_L - \epsilon_0}{\Gamma/2}} \right )  \nonumber \\ 
&+\frac{1}{\pi \omega}\left(\frac{\Gamma/2}{\omega^2 + \Gamma^2}\right)
\ln{\frac{(\mu_L-\epsilon_0)^2 + \frac{\Gamma^2}{4}}{(\mu_L + \omega - \epsilon_0)^2 + \frac{\Gamma^2}{4}}}
\label{t2_zeroT}
\end{eqnarray}

\section{High-T limit of phonon rate equations}

In this section we shall show how for weak electron phonon coupling and
at high-T the quantum Boltzmann equation for phonons (Eq.~\ref{qb_ph1}, ~\ref{R_in}, ~\ref{R_out}) 
reduces to the classical rate equations derived in Section II. For $T \gg \Gamma$, we can
replace the lorentzian broadenings in Eqn~\ref{R_in} and \ref{R_out} by $\delta$ functions
{\it {i.e.}} ($\frac{\Gamma/2}{x^2 + \Gamma^2/4} \rightarrow \pi \delta(x)$). Moreover for 
weak electron-phonon couplings we may identify the probability of single occupancy
of the dot to be given by its value in the absence of electron phonon coupling,
{\it {i.e.}} $P^1 = \frac{\sum_a\Gamma_a f(\epsilon_0 - \mu_a)}{\Gamma}$, so that 
\begin{eqnarray}
&R_{n \rightarrow n+1} = (n+1) \lambda^2 \frac{1}{1 + \frac{\Gamma^2}{4\omega_0^2}}\{ P^1 
\sum_b \Gamma_b (1 - f(\epsilon_0 - \omega_0 - \mu_b)) 
\label{R1qcl} \nonumber \\ 
& +  P^0 
\sum_b \Gamma_b f(\epsilon_0 + \omega_0 - \mu_b)  \} 
\end{eqnarray}
and 
\begin{eqnarray}
&R_{n \rightarrow n-1} = n \lambda^2 \frac{1}{1 + \frac{\Gamma^2}{4\omega_0^2}}\{ P^1 
\sum_b \Gamma_b (1 - f(\epsilon_0 + \omega_0 - \mu_b)) 
\label{R2qcl} \nonumber \\ 
& +  P^0 
\sum_b \Gamma_b f(\epsilon_0 - \omega_0 - \mu_b)  \} 
\end{eqnarray}

Now the classical rate equations in Eq.~\ref{rateeqs} rewritten for the case of a single
resonant level with $U=0$ take the form,
\begin{eqnarray}
\dot{P}^0_n &=& - P^0_n \sum_{a,m} R^{a,0,1}_{n,m} + \sum_{a,m} P^1_m R^{a,1,0}_{m,n}
\label{p00}\\
\dot{P}^1_n &=& - P^1_n \sum_{a,m} R^{a,1,0}_{n,m} + \sum_{a,m} P^0_m R^{a,0,1}_{m,n}
\label{p11}
\end{eqnarray}
where $R^{a,0,1}_{n,m}$ etc are defined in Eqns.~\ref{Rdef1} and \ref{Rdef2} with $U=0$.
In the weak electron-phonon coupling limit, only  transitions that change the
phonon number at most by 1 are allowed. In this limit the 
$\Gamma^a_{n,n+1} \rightarrow (n+1) \lambda^2 \Gamma_a$. Moreover we may again factorize
the joint probability distribution of having $0/1$ electrons and $m$ phonons as 
$P^{0,1}_m = P^{0,1} P^{ph}_m$,  
After making these approximations in Eq.~\ref{p00} and \ref{p11}
and adding the two equations we obtain 
\begin{eqnarray}
\dot{P}^{ph}_n &=& -P^{ph}_n \left (R_{n \rightarrow n+1} + R_{n \rightarrow n-1} \right)  
\\ \nonumber
&+& P^{ph}_{n+1} R_{n+1 \rightarrow n} +  P^{ph}_{n-1} R_{n-1 \rightarrow n}
\end{eqnarray}
with the rates given by Eq.~\ref{R1qcl} and \ref{R2qcl}.



\begin{thebibliography}{200}
\bibitem{Organicdevice}
J. Chen, M. A. Reed, A. M. Rawlett, and J. M. Tour, \textsl{Science},
\textbf{286},1550 (1999);
C.P. Collier, G. Mattersteig, E. W. Wong, Y. Luo, K. Beverly, J, Sampaio, F. M.
Raymo, J. F. Stoddart, and J. R. Heath, \textsl{Science},\textbf{289}, 1172
(2000); 
W. Liang, M. P. Shores, M. Brockrath, J. R. Long and
H. Park, \textsl{Nature}, \textbf{417}, 725 (2002); 
J. Reichert, R. Ochs, D. Beckmann, H. B. Weber, M. Mayor, and H. v. 
Loheysen, \textsl{Phys. Rev. Lett.}, \textbf{88}, 176804 (2002);
 N. B. Zhitenev, H. Meng, and Z. Bao, \textsl{Phys. Rev. Lett.},
\textbf{88}, 226801, (2002); 
S. Kubatkin, A. Danilov, M. Hjort, J. Cornil, J. Bredas, 
N. Stuhr-Hansen, P. Hedegard and T. Bjornholm, {\sl{Nature}},{\bf{425}},68 (2003).

\bibitem{Quantumdotrefs}L. P. Kouwenhoven and C. M. Marcus, Phys. World
\textbf{11} (6) , 35 (1998)and references therein. Also L. P. Kouwenhoven, C. M.
Marcus, P.L. McEuen, S. Tarucha, R. M. Westervelt, and N. S. Wingreen Electron
Transport in Quantum Dots Nato ASI conference proceedings, Series E345, ed. By L. P.
Kouwenhoven, G. Schon, L.L. Sohn (Kluwer, Dordrecht, 1997).

\bibitem{Quantumdottheory}see, e.g. \textsl{Quantum effects in Coulomb
blockade},I. L Aleiner, P. W. Brouwer and L. I Glazman, \textsl{Physics
Reports}, \textbf{358}, 309 (2002) and references therein.

\bibitem{Ralph02}J. Park, A. N. Pasupathy, J. I. Goldsmith, C. Chang, Y.
Yaish, J. R. Petta, M. Rinkoski, J. P. Sethna, H Abruna, P. L. McEuen, and D.
C. Ralph,\textsl{Nature}, \textbf{17}, 722 (2002).


\bibitem{Park00}H. Park, J. Park, A. K. L. Lim, E. H. Anderson, A. P.
Alivisatos, and P. L. McEuen, \textsl{Nature}, \textbf{407}, 57 (2000).

\bibitem{Ralph03} A. N. Pasupathy, J. Clark, C. Chang, A. V. Soldatov, S. Lebedkin,
R. C. Bialczak, J. E. Grose, L. A. K. Donev, J. P. Sethna, D. C. Ralph and
P. L. McEuen, cond-mat/0311150.

\bibitem{Glazman87}L. Glazman and R. I. Shekhter, Zh. Eksp. Teor. Fiz.
\textbf{94 }292 (1987) [Sov. Phys. JETP \textbf{67 }163 (1988)].

\bibitem{Wingreen89}N. S. Wingreen, K. W. Jacobsen, and J. W. Wilkins,
\textsl{Phys. Rev. B.}, \textbf{40}, 11834 (1989).

\bibitem{McKenzie} U. Lundin and  H. McKenzie, Phys. Rev. B., {\bf{66}}, 075303 (2002).

\bibitem{Balatsky02}J. X. Zhu and A. V. Balatsky, Phys. Rev. B {\bf{67}}, 
165326 (2003).

\bibitem{Prokofiev03} K. D. McCarthy, N. Prokofev, and M. T. Tuominen, 
\textsl{Phys. Rev. B}, \textbf{67}, 245415 (2003).

\bibitem{Aji} V. Aji, J. E. Moore, and C. M. Varma, cond-mat/0302222.

\bibitem{Gogolin02}A. O. Gogolin and A. Komnik, unpublished (cond-mat/0207513).

\bibitem{Alexandrov02}A. S. Alexandrov and A. M. Bratkovsky, and R. S. Williams,
Phys. Rev. B. {\bf{67}}, 075301 (2003); S. Alexandrov and A. M. Bratkovsky,
Phys. Rev. B. {\bf{67}}, 235312 (2003).   

\bibitem{Flensberg} Karsten Flensberg, \textsl{Phys. Rev. B}, \textbf{68},
205323 (2003). 

\bibitem{Fedorets} D. Fedorets, cond-mat/0311104; D. Fedorets, L. Y. Gorelik, R. I. Shekhter, and 
M. Jonson, cond-mat/0311105; D. Fedorets, L. Y. Gorelik, R. I. Shekhter, and 
M. Jonson, Europhys. Let. {\bf{58}},  99 (2002).

\bibitem{Zoller} C. W. Gardiner and P. Zoller, {\sl{ Quantum Noise}}, Springer
Verlag, 2nd Edition.  

\bibitem{Wingreen94} A. Jauho, N. S. Wingreen, and Y. Meir, Phys. Rev. B. {\bf{50}}, 
5528 (1994).

\bibitem{Rate_Eqn}See, for example L. I. Glazman and K. Matveev,
Pis'ma Zh. Eksp. Theor. Fiz. {\bf 48} 403 (1988)
or C. W. J. Beenakker, \textsl{Phys. Rev. B},
\textbf{44},1646 (1991).

\bibitem{Mahanch4} See, e.g. section 4.3 of G. D. Mahan, \textit{Many
Particle Physics}, 2nd Edition, Plenum Publishers.

\bibitem{Korotkov} A. N. Korotkov, Phys. Rev. B {\bf{49}}, 10381 (1994). 

\bibitem{Keldysh_ref} L. V. Keldysh, Zh. Eksp. Teor. Fiz.,
{\bf{47}}, 1945 (1964) [Sov. Phys. JETP {\bf{46}}, 155 (1977)].

\bibitem{Vavilov_Aleiner} M. G. Vavilov and I. L. Aleiner, \textsl{Phys. Rev. B},
{\bf 69}, 035303  (2004). 

\bibitem{Schon94} H. Schoeller, and G. Sch{\o}n, \textsl{Phys. Rev. B}, 
{\bf 50}, 18436 (1994).

\bibitem{Rammer} J. Rammer, Rev. Mod. Phys. {\bf{63}}, 781 (1991). 

\bibitem{Flensberg03} S. Braig and K. Flensberg, cond-mat/0401347.

\end{thebibliography}
\end{document}